\documentclass[conference]{IEEEtran}
\IEEEoverridecommandlockouts
\usepackage{cite}
\usepackage{amsmath,amssymb,amsfonts,pifont}
\usepackage{algorithm}
\usepackage{algorithmic}
\usepackage{hyperref}
\usepackage{graphicx}
\usepackage[misc]{ifsym}
\usepackage{diagbox}
\usepackage{textcomp}
\usepackage[dvipsnames, svgnames, x11names]{xcolor}
\usepackage{bigstrut}
\usepackage{url}
\usepackage{booktabs}
\usepackage{multirow}
\usepackage{microtype}
\usepackage{subfigure}
\usepackage{makecell}
\usepackage{adjustbox} % For adjusting table size
\usepackage{enumitem}
\usepackage{listings}
\usepackage{wrapfig}
\usepackage{caption}
\usepackage{bm}
\usepackage{color}
\usepackage{hyphenat}
\usepackage{balance}
\usepackage[most]{tcolorbox}
\tcbuselibrary{skins, breakable, theorems}
\usepackage{mathrsfs}
\usepackage{mathtools}
\usepackage{xspace}
\usepackage{bbding}

\makeatletter
\newcommand{\removelatexerror}{\let\@latex@error\@gobble}
\newcommand{\linebreakand}{
\end{@IEEEauthorhalign}
\hfill\mbox{}\par\mbox{}\hfill
\begin{@IEEEauthorhalign}
}
\makeatother

\definecolor{codegreen}{rgb}{0,0.3,0.6}
\definecolor{codegray}{rgb}{0.5,0.5,0.5}

\newcommand{\ie}{\emph{i.e.,}\xspace}
\newcommand{\eg}{\emph{e.g.,}\xspace}

\newcommand{\ignore}[1]{}

\def\BibTeX{{\rm B\kern-.05em{\sc i\kern-.025em b}\kern-.08em
    T\kern-.1667em\lower.7ex\hbox{E}\kern-.125emX}}

\begin{document}

\title{Adapting Large Language Models by Integrating Collaborative Semantics for Recommendation

\thanks{\Letter\ Corresponding author.}
}

\author{\IEEEauthorblockN{Bowen Zheng\IEEEauthorrefmark{1},
Yupeng Hou\IEEEauthorrefmark{2}, Hongyu Lu\IEEEauthorrefmark{3}, Yu Chen\IEEEauthorrefmark{3},
Wayne Xin Zhao\IEEEauthorrefmark{1}\textsuperscript{\Letter}, Ming Chen\IEEEauthorrefmark{3}, and Ji-Rong Wen\IEEEauthorrefmark{1}}
\IEEEauthorblockA{\IEEEauthorrefmark{1}Gaoling School of Artificial Intelligence,
Renmin University of China, China\\
\IEEEauthorrefmark{2}University of California San Diego, United States\\
\IEEEauthorrefmark{3}WeChat, Tencent, China\\
bwzheng0324@ruc.edu.cn, yphou@ucsd.edu, luhy94@gmail.com, nealcui@tencent.com, \\ batmanfly@gmail.com, mingchen@tencent.com, jrwen@ruc.edu.cn}}

\maketitle

\begin{abstract}
Recently, large language models~(LLMs) have shown great potential in recommender systems, either improving existing recommendation models or serving as the backbone. However, there exists a large semantic gap between LLMs and recommender systems, since items to be recommended are often indexed by discrete identifiers (\emph{item ID}) out of the LLM's vocabulary. In essence, LLMs capture language semantics while recommender systems imply collaborative semantics, making it difficult to sufficiently leverage the model capacity of LLMs for recommendation.  

To address this challenge, in this paper, we propose a new LLM-based recommendation model called \emph{LC-Rec}, which can better integrate language and collaborative semantics for recommender systems. 
Our approach can directly generate items from the entire item set for recommendation, without relying on candidate items.
Specifically, we make two major contributions in our approach. For item indexing, we design a learning-based vector quantization method with uniform semantic mapping, which can assign meaningful and non-conflicting IDs (called \emph{item indices}) for items. For alignment tuning, we propose a series of specially designed tuning tasks to enhance the integration of collaborative semantics in LLMs. Our fine-tuning tasks enforce LLMs to deeply integrate language and collaborative semantics (characterized by the learned item indices), so as to achieve an effective adaptation to recommender systems. 
Extensive experiments demonstrate the effectiveness of our method, showing that our approach can outperform a number of competitive baselines including traditional recommenders and existing LLM-based recommenders. 
Our code is available at \url{https://github.com/RUCAIBox/LC-Rec/}.
\end{abstract}


\section{Introduction}
\label{sec:intro}

Nowadays, recommender systems have become an essential part of various application platforms, aiming to recommend potential information resources to users based on their specific preferences. Since user preferences dynamically evolve over time, \emph{sequential recommendation} has attracted great research attention due to its advantages in 
capturing the sequential characteristics of user behaviors. To develop sequential recommenders, existing recommendation models~\cite{hidasi2015session,kang2018self} are mostly built on sequential formatting of user interaction logs, taking \emph{item ID} as the basic unit. 
%Sequential recommendation, in particular, has attracted great interest due to its advantages in capturing the sequential characteristics of user behaviors.

In existing literature, sequential recommenders adopt various deep neural networks to model user historical behaviors represented by item ID sequences, including RNN~\cite{hidasi2015session,li2017neural}, CNN~\cite{tang2018personalized,yuan2019simple}, GNN~\cite{wu2019session,xu2019graph}, and Transformer~\cite{kang2018self,sun2019bert4rec}.
In addition to collaborative semantics within users' historical behaviors, some studies~\cite{zhang2019feature,zhou2020s3,xie2022decoupled} also try to enhance item sequence modeling by leveraging content information (\eg title, description, category).
Moreover, pre-trained language models~(PLMs) have also been employed for capturing the textual semantics reflected in item texts~\cite{hou2022towards,li2023text,ding2021zero,hou2023learning}, to improve the recommendation performance.  

%the language information in item titles and descriptions has motivated many works to explore the utilization of Pre-trained Language Models (PLMs) in RS~\cite{hou2022towards,li2023text,ding2021zero,hou2023learning}.

Recently, the emergence of large language models (LLMs) has triggered a significant revolution in the research community. 
LLMs have shown great potential in various language based tasks, due to their excellent capabilities in semantic understanding and generation~\cite{zhao2023survey}. Specifically, there are also several attempts~\cite{wu2023survey} that adapt LLMs for recommender systems~(RS), to improve the item ranking performance~\cite{hou2023large} or boost the comprehensive recommendation capacities~\cite{zhang2023recommendation}. 
To develop capable LLM-based recommendation models, a fundamental challenge is that there exists a large gap between the \emph{language semantics} modeled by LLMs and \emph{collaborative semantics} implied by recommender systems.
The key point is that, in existing recommendation models, user behaviors are often formatted into item ID sequences (possibly with feature IDs), but not textual descriptions. 
In other words, language models and recommendation models indeed employ two different vocabularies (token IDs \emph{v.s.} item IDs) to learn their own semantic spaces. Such a semantic gap makes it difficult to sufficiently leverage the model capacity of LLMs for tackling the recommendation tasks.

To address this issue, existing efforts can be divided into two main approaches. The first approach~\cite{hou2023large,cui2022m6,zhang2023recommendation,bao2023tallrec} verbalizes the user behaviors into text sequences (\eg concatenating the titles and category labels of the interacted items), and designs special prompts to instruct LLMs for fulfilling the recommendation tasks. Such an approach only captures limited item information (only considering language semantics),  and can't guarantee the generation of in-domain items (relying on a candidate set). As another alternative approach, several studies~\cite{geng2022recommendation,hua2023index} design special item indexing mechanisms for building item vocabulary, and then learn to generate the target item for recommendation.
%A straightforward implementation is to extend the text vocabulary by including item IDs and then fine-tune LLMs via the target recommendation task. 
However, given the large semantic gap,
\emph{simple} (\eg vocabulary building with vanilla item IDs) or \emph{shadow} integration (\eg fine-tuning only with the target task) would be less effective to adapt LLMs for recommender systems. 

Considering these issues, we aim to design a more effective semantic integration approach for developing LLM-based recommendation models. We tackle this semantic integration problem in two main aspects, namely item indexing and alignment tuning\footnote{Note that the words ``align'' and ``alignment'' mainly refer to the integration between language semantics and collaborative semantics, but not what it means in human alignment~\cite{ouyang2022training} that instructs LLMs to follow human values or preferences.}.  
For item indexing, 
an ideal allocation mechanism should produce \emph{meaningful} (capturing item similarities), \emph{unique} (without allocation conflicts), and \emph{extensible} (generalizable to new items) IDs for effectively representing the items.    
%should be able to reflect the similarity or relevance among items (\eg textual or collaborative similarities), meanwhile ensuring no conflicts and can generalize to new items (\ie avoiding the out-of-vocabulary~(OOV) issue). 
For alignment tuning, it should be able to sufficiently integrate language semantics with collaborative semantics in LLMs, but not superficially fit the target recommendation task. Overall, our goal is to effectively establish the connections between the two kinds of different semantics and fully leverage the model capacity of LLMs for sequential recommendation.

To this end, in this paper, we propose \textbf{LC-Rec}, a new approach to integrate \textbf{L}anguage and \textbf{C}ollaborative semantics for improving LLMs in \textbf{Rec}ommender systems.
Our approach is built in a generative manner, where the recommendation task is cast into a token generation task as well. 
To achieve this, the key point lies in the semantic integration between language and collaborative semantics, so that LLMs can make the item recommendations just like they generate normal text contents.
Our approach has made two major contributions in the aforementioned two aspects. 
For item indexing, we propose a tree-structured vector quantization~(VQ) method to index the items with discrete IDs (called \emph{item indices}). 
These item indices are learned based on the text embeddings of items encoded by LLMs, enabling the learned IDs to capture the intrinsic similarity among items. 
However, original VQ methods are likely to assign the same IDs to multiple items, which should be avoided in recommender systems. To tackle this problem, we further design a uniform semantic mapping method to mitigate the potential conflicts in ID allocation. 
For alignment tuning, we design a series of specific tasks to fine-tune LLMs for achieving semantic integration. In addition to the sequential item prediction, we consider both \emph{explicit index-language alignment} and \emph{implicit recommendation-oriented alignment}. Our fine-tuning tasks enforce LLMs to deeply integrate language and collaborative semantics, so as to achieve an effective adaptation to recommender systems. 

To evaluate our approach, we conduct extensive experiments on three real-world datasets. Our method achieves the best performance compared to a number of competitive baselines. 
Experimental results demonstrate that our approach can effectively align language and collaborative semantics via specially learned item indices, thereby significantly improving the recommendation performance. The contributions of this work can be summarized as follows:
\begin{itemize}
\item We present LC-Rec, a LLM-based sequential recommendation model, by effectively integrating language and collaborative semantics. LC-Rec can fulfill the sequential recommendation task in an autoregressive generation way, without relying on candidate sets.  
%which more effectively utilizes the semantic comprehension and generation capabilities of LLMs, thereby achieving better full ranking in sequential recommendation.
\item Our approach is built on a specially designed VQ method, which can capture item similarity and avoid ID conflicts in index allocation. Further, we propose a series of carefully designed tuning tasks for achieving effective semantic integration via item indices.
%We design a uniform semantic mapping method to effectively resolve conflicts during item index construction. Additionally, we seamlessly integrate items among recommendation scenarios into LLM through a series of carefully designed semantic-related tasks.
\item We implement our method based on LLaMA~\cite{touvron2023llama} with 7B parameters. Extensive experiments on three public datasets demonstrate the effectiveness of our approach in integrating collaborative semantics into LLMs. The proposed method LC-Rec achieves an average performance improvement of 25.5\% in full ranking evaluations, compared to all baseline methods.
\end{itemize}

\section{Related Work}

\subsection{Sequential Recommendation}
Sequential recommendation aims to infer user preferences by analyzing historical interactions and predict the next item that would be suitable for that user~\cite{hidasi2015session,sun2019bert4rec,kang2018self}.
Many early methods are frequently based on Markov Chains techniques~\cite{rendle2010factorizing,he2016fusing}.
Recently, 
% due to the rapid advancement of deep neural networks, 
typical methods become to adopt various deep neural networks to model user historical behaviors represented by item ID sequences, including RNN~\cite{hidasi2015session,li2017neural}, CNN~\cite{tang2018personalized,yuan2019simple}, GNN~\cite{wu2019session,xu2019graph}, and Transformer~\cite{kang2018self,sun2019bert4rec}.
However, these methods only capture the collaborative relationship between items from user-item interactions, while ignoring the additional information rich in the item content information (\eg title, description, category).
Therefore, several studies are devoted to utilizing additional information associated with items to enhance ID sequence modeling~\cite{zhang2019feature,zhou2020s3,xie2022decoupled}.
% Some of them use content information as auxiliary supervision signals~\cite{zhou2020s3}, while others focus on learning content-related embeddings to extend item representation. 
Furthermore, the inherent natural language characteristics of the item title and description have motivated numerous researchers to explore the utilization of pre-trained language models (PLMs) in recommender systems~\cite{hou2022towards,li2023text,ding2021zero,hou2023learning}.
In this paper, we aim to combine LLMs and recommendation tasks in a more effective way, which is reached through the proposed new item indexing and alignment tuning methods.

% hope to integrate the powerful semantic understanding capabilities of language models into sequential recommendation and achieve more excellent performance.

\begin{figure*}[]
\centering
\includegraphics[width=0.98\linewidth]{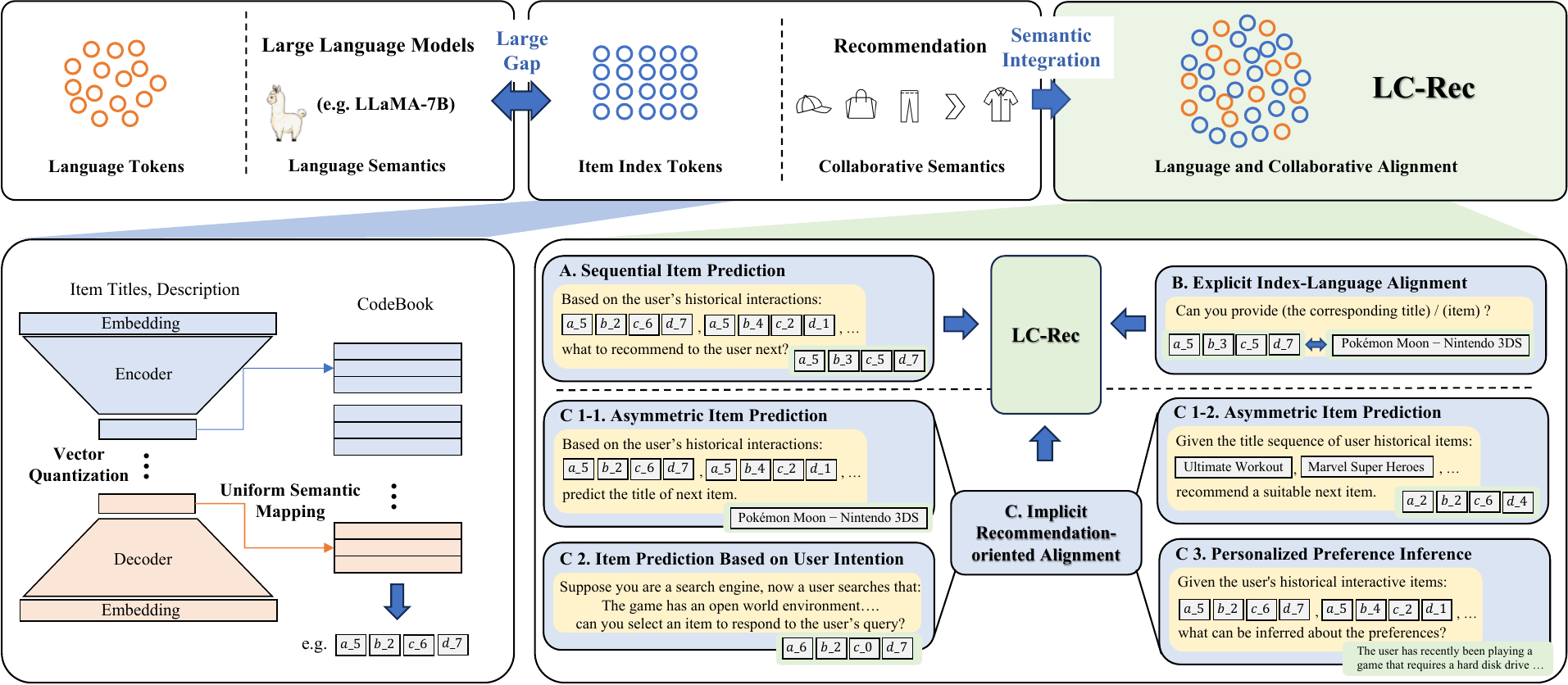}
\caption{The overall framework of our LC-Rec. We enhance language learning models (LLMs) by integrating language and collaborative semantics based on item indexing and alignment tuning, thereby adapting LLMs to recommender systems.}
\label{fig:model}
\end{figure*}

\subsection{Large Language Models for Recommendation}

Recently, large language models (LLMs) have gained significant popularity, with a wide range of applications spanning various domains of artificial intelligence~\cite{zhao2023survey,yao2023react,zhang2023speechgpt,jiang2023structgpt}.
This is largely attributed to their superior capabilities in language semantic understanding and generation.
In the context of RS, researchers have been working on adapting LLMs for RS to improve recommendation performance. A common approach is to represent user behaviors as text sequences (\eg by concatenating the titles of historical items), and then design prompts to guide LLMs to perform the recommendation task~\cite{gao2023chat,dai2023uncovering,wang2023zero}.

% Furthermore, some studies~\cite{hou2023large,liu2023chatgpt,wang2023rethinking} improve recommendation performance by emphasizing the recent interactive item or applying in-context learning (ICL).

However, a major challenge remains: there is a large gap between the language semantics modeled by LLMs and the collaborative semantics implied by recommender systems, which cannot be bridged by simple prompt design alone. To address this problem, existing efforts can be categorized into two main approaches. 
The first approach is to fine-tune the LLMs with text-based user behavior sequences~\cite{cui2022m6,zhang2023recommendation,bao2023tallrec,yue2023llamarec}. However, these methods cannot guarantee the generation of in-domain items. Due to the constraints of limited context window size, these methods can only rank on a given candidate set, and can hardly be applied in a full ranking scenario.
% adapted for ranking scenarios. % can only capture a limited amount of item information and may miss collaborative semantic information.  
% In addition, text-based methods mainly consider language semantics, potentially overlooking vital collaborative semantics.
The second approach maintains the use of item IDs or introduces unique item indexing mechanisms.
Given pure item ID sequences, models are trained to directly generate target item IDs for recommendation~\cite{geng2022recommendation,hua2023index}. 
% Nevertheless, the input of the recommendation task is basically pure item indices.  establishes
Although collaborative semantics between item indices are established, the language semantics modeled by LLMs and these item indices are not well aligned.
% shadow semantic integration}
% ignores the alignment with language semantics modeled by LLMs.}

Additionally, we are aware of some concurrent works~\cite{lin2023multifacet,zhang2023collm,zhu2023collaborative}, which also recognize the issue of the large semantic gap between recommendation tasks and natural language tasks. For instance, TransRec~\cite{lin2023multifacet} employs multi-facet identifiers,  combining ID, title, and attributes to balance item distinctiveness and semantics.
% that incorporate ID, title, and attribute to achieve both item distinctiveness and semantics; 
CoLLM~\cite{zhang2023collm} incorporates collaborative semantics into LLMs by using representations of an external collaborative model as part of the input. CLLM4Rec~\cite{zhu2023collaborative} learns dual user/item embeddings based on recommendation task and content (\eg reviews) generation task, respectively. A mutually-regularization loss is introduced for interaction between these two kinds of embeddings. 
Only the recommendation-task embeddings are for the final recommendation.
% In contrast to these approaches, which tend to model the different aspects of item semantics separately (\eg multiple identifiers~\cite{lin2023multifacet}, external representations~\cite{zhang2023collm}, two types of embedding~\cite{zhu2023collaborative}), our focus is on enabling deep and unified language and collaborative semantic integration in LLMs via specially learned item indices.

Contrasting these methodologies, our focus is to develop a deep and unified integration of language and collaborative semantics within LLMs through carefully crafted item indices.
In particular, our approach uses a tree-structured vector quantization method to construct item indices. 
% This method allows us to minimize the difficulty of semantic integration by adding only a small number of additional tokens to LLMs. 
This method allows for better semantic integration by adding a small number (usually $\sim$1,000) of additional tokens to LLMs.
Furthermore, we introduce a series of semantic alignment tasks to fine-tune LLMs, aiming to achieve unified semantic integration in a practical recommendation setting (\eg full ranking).

\section{Methodology}
% In this section, we present the proposed LLM-based recommendation method  LC-Rec, which aligns language and collaborative semantics with item indices. This way, we achieve the purpose of integrating items into the LLM and adapting it for sequential recommendation tasks.
In this section, we present the proposed LLM-based recommendation model \textbf{LC-Rec}, which integrates \textbf{L}anguage and \textbf{C}ollaborative semantics for improving LLMs in \textbf{Rec}ommender systems.

\subsection{Overview of the Approach}

As we discussed in Section~\ref{sec:intro}, there exists a large gap between the \emph{language semantics} modeled by LLMs and \emph{collaborative semantics} implied by recommender systems, which limits the capacities of LLMs in recommender systems.
To effectively bridge this gap, we consider enhancing the semantic integration in two major aspects.

\begin{itemize}
\item For \emph{item indexing}  (Section~\ref{sec:index}), we represent an item with several learned discrete IDs via vector quantization based on text embeddings by LLMs, and further propose a uniform semantic mapping method to mitigate the potential conflicts in index assignment.
In this way, the learned item indices can capture similarities between the textual semantics of item information, and provide a unique indexing representation for a specific item. 
\item For \emph{alignment tuning} (Section~\ref{sec:tuning}), we design a series of specific tuning tasks that enhance the integration between language semantics and collaborative semantics, not limited to the target recommendation task. Our approach can effectively integrate the collaborative semantics into LLMs, and sufficiently leverage the powerful model capacity of LLMs for recommendation tasks. 
\end{itemize}

The overall framework of the proposed approach LC-Rec is shown in Figure~\ref{fig:model}. Next, we will present the details of our method.

\subsection{Learning Item Indices for Semantic Integration}
\label{sec:index}

To extend the capacities of LLMs for recommendation, a fundamental problem is how to represent an item with index IDs (called \emph{item indices}) and integrate these item indices into LLMs. We don't adopt the original item ID (resulting in a very large vocabulary), but instead employ vector quantization techniques to represent an item with a small number of discrete indices. 
These indices are constructed by leveraging relevant item information (\eg item text representations), and the token embeddings associated with these discrete indices can be further optimized to fit the recommendation task (Section~\ref{sec:tuning}). In this part, we present the approach for learning item indices for subsequent semantic integration.
The approach consists of two major steps: it first conducts vector quantization based on text embeddings of items, so that the original representations of item indices can capture latent textual semantic correlations between items; then, it proposes a uniform semantic mapping to mitigate the potential conflicts in item index assignment. 
Next, we introduce the two parts in detail. 

\subsubsection{Vector Quantization Indexing} 
In recommender systems, it is common to associate each item with a single unique ID (called vanilla ID).
However, it would directly introduce a large vocabulary of item IDs when dealing with a great number of items (\ie a large item set). Further, such an approach is easy to suffer from the OOV issue when adapting to new items (\eg cold-start items). To address this issue, we borrow the idea of existing studies~\cite{hou2023learning,rajput2023recommender,hua2023index} to learn indices associated with latent semantics for items. Specifically, each item is represented by a composition of discrete indices corresponding to its own latent semantic, and each discrete index can be shared by multiple items. 
The basic idea is that similar items tend to be assigned with a portion of common semantic indices, such that each unique semantic index can be aligned to some kind of latent semantics.  

To derive these semantic indices, we first employ LLMs (\eg LLaMA) to encode the attached text information for an item, and obtain the text embeddings as the initial item representation. %In this way, the initial item representations encode textual semantics, making that similar items have similar initial representations. 
Further, we propose to use a Vector Quantization~(VQ) approach to create discrete indices based on item embeddings. 
%Naturally, the title and description associated with each item are our best source of preliminary semantic information. Following this, drawing inspiration from Vector Quantization (VQ) technology, we employ a method that combines item embedding and VQ to create discrete indices for items.
Specifically, we take the item embeddings encoded by LLMs as input, and then train a Residual-Quantized Variational AutoEncoder (RQ-VAE) for generating item indices. 
%we first use the Language Model (LLM) to encode titles and descriptions of all items. Then, the encoded item embeddings are used as input to train a Residual-Quantized Variational AutoEncoder (RQ-VAE) for generating item indices.
RQ-VAE~\cite{zeghidour2021soundstream} is a multi-level vector quantizer, which recursively quantized the residual vectors from coarse to fine to generate a set of codewords (\ie item indices). 
For an item embedding $\bm{e}$, RQ-VAE first encodes it into a latent representation $\bm{z}$. At each level $h$, we have a codebook $\mathcal{C}^h = \{\bm{v}_k^h\}_{k=1}^K$, where each codebook vector $\bm{v}_k^h$ is a learnable cluster center. Then the residual quantization process can be expressed as:
\begin{align}
    & c_i = \underset {k} { \operatorname {arg\,min} } ||\bm{r}_i - \bm{v}_k^i||_2^2, 
    \label{eq:rqc} \\
    & \bm{r}_{i+1} = \bm{r}_i - \bm{v}_{c_i}^i,
    \label{eq:rqr}
\end{align}
where $c_i$ is the $i$-th codeword of the item indices and $\bm{r}_i$ is the residual vector in the $i$-th RQ level, and we set $\bm{r}_1 = \bm{z}$.

When we have $H$-level codebooks, the quantization representation of $\bm{z}$ can be obtained according to $\hat{\bm{z}} = \sum_{i=1}^{H} \bm{v}_{c_i}^i$. 
Then $\hat{\bm{z}}$ will be used as decoder input to reconstruct the item embedding $\bm{e}$.
The overall loss function is as follows:
\begin{align}
    \mathcal{L}_{\text{RECON}} &= ||\bm{e} - \hat{\bm{e}}||_2^2, \\
    \mathcal{L}_{\text{RQ}} &= \sum_{i=1}^{H} ||\text{sg}[\bm{r}_i] - \bm{v}_{c_i}^i||_2^2 + \beta \ ||\bm{r}_i - \text{sg}[\bm{v}_{c_i}^i]||_2^2, \\
    \mathcal{L}_{\text{RQ-VAE}} & = \mathcal{L}_{\text{RECON}} + \mathcal{L}_{\text{RQ}},
    \label{eq:rqloss}
\end{align}
where $\hat{\bm{e}}$ is the output of the decoder, $\text{sg}[·]$ represents the stop-gradient operator, and $\beta$ is a loss coefficient, usually set to 0.25. 
The overall loss is divided into two parts, $\mathcal{L}_{\text{RECON}}$ is the reconstruction loss, and $\mathcal{L}_{\text{RQ}}$ is the RQ loss used to minimize the distance between codebook vectors and residual vectors.

Compared with traditional VQ approaches, RQ offers the advantage of achieving a larger expression space with a smaller codebook size~\cite{zeghidour2021soundstream,lee2022autoregressive}.
% \textcolor{blue}{Under the condition of equal overall codebook size~(\eg $H$-level codebooks and the size of each level codebook is $K$), the expression space of RQ~(\ie $K^H$) is much larger than that of traditional VQ methods~(\ie $KH$)~\cite{zeghidour2021soundstream,lee2022autoregressive}}.
Besides, its coarse-to-fine quantification method results in a tree-structured item index, which is beneficial for autoregressive generation.
In fact, the RQ approach has demonstrated its effectiveness across various autoregressive generation tasks, such as autoregressive image generation~\cite{lee2022autoregressive} and generative recommendation~\cite{rajput2023recommender}.
Instead of simply employing VQ for item indexing~\cite{hou2023learning,rajput2023recommender}, we consider two key improvements for deriving meaningful item indices. First, there should be no conflicts in item indices, which is a common issue with VQ but should not occur in recommender systems. Second, the established semantic spaces of item indices should be aligned with the semantics of LLMs, in order to better leverage the powerful model capacity of LLMs for recommendation. We next introduce the two major improvements in our approach.

\begin{algorithm}[t]
\centering
\caption{RQ with Uniform Semantic Mapping}
\label{alg:rq-usm}
\begin{algorithmic}[1]
\REQUIRE Batch item representations $\bm{B} = \{\bm{z}^n\}_{n=1}^{|\bm{B}|}$; $H$-level codebooks $\{\mathcal{C}^h\}_{h=1}^{H}$.
\ENSURE  Item indices $\{ [c_1^n, c_2^n,... ,c_H^n] \}_{n=1}^{|\bm{B}|}$; Quantified representations $\{\hat{\bm{z}}^n\}_{n=1}^{|\bm{B}|}$.
\STATE Let initial residual vectors $\bm{r}_1^n = \bm{z}^n, \ \forall \bm{z}^n \in \bm{B}$
\FOR{$i = 1 $ \TO $H$}
    \IF{$i<H$}
        \STATE Solve $\{c_i^n\}_{n=1}^{|\bm{B}|}$ according to Eqn.~\eqref{eq:rqc}
    \ELSE
        \STATE Solve $\{c_H^n\}_{n=1}^{|\bm{B}|}$ according to Eqn.~\eqref{eq:rq-skh} via Sinkhorn-Knopp algorithm
    \ENDIF
    \STATE Obtain $\{\bm{r}_{i+1}^n\}_{n=1}^{|\bm{B}|}$ according to Eqn.~\eqref{eq:rqr}
\ENDFOR
\FORALL{$\bm{z}^n \in \bm{B}$}
    \STATE Calculate quantified representations by $\hat{\bm{z}}^n = \sum_{i=1}^{H} \bm{v}_{c_i^n}^i$
\ENDFOR
\RETURN $\{ [c_1^n, c_2^n,... ,c_H^n] \}_{n=1}^{|\bm{B}|}$ and $\{\hat{\bm{z}}^n\}_{n=1}^{|\bm{B}|}$
\end{algorithmic}
\end{algorithm}

\subsubsection{Conflict Mitigation via Uniform Semantic Mapping} 
Since we adopt the tree structure for learning item indices, it might lead to index conflicts among items within the same leaf node. 
To address this issue, existing solutions~\cite{rajput2023recommender,hua2023index} typically add an additional layer to the index tree and assign a 
% unique extra 
distinct supplementary index ID to each item in a node with conflicts. However, this approach introduces semantically irrelevant distributions 
% at the last layer of the index tree
in the tree's final layer. Additionally, these newly integrated IDs might also affect the original item representations.

Considering these issues, we propose a new conflict mitigation method to avoid the clustering of multiple items within the same leaf node.
%our approach focuses on avoiding the clustering of multiple items within the same leaf node. 
Our objective is to ensure that item semantics are uniformly distributed across different codebook embeddings at the last index level. To achieve this, we introduce a uniform distribution constraint to the original formulation:
\begin{align}
\begin{split}
    \min & \sum_{\bm{r}_{H} \in \bm{B}} \sum_{k=1}^K q(c_{H}=k|\bm{r}_{H}) ||\bm{r}_{H} - \bm{v}_k^H||_2^2, \\
    \text{subject \ to}:
    & \ \sum_{k=1}^K \ q(c_{H}=k|\bm{r}_{H}) = 1, \\
    & \sum_{\bm{r}_{H} \in \bm{B}} q(c_{H}=k|\bm{r}_{H})  = \frac{|\bm{B}|}{K},
    \label{eq:rq-skh}
\end{split}
\end{align}
where $\bm{B}$ is a batch of residual vectors in the last index level. 
Following~\cite{zhan2022learning,asano2019self,caron2020unsupervised,lin2022prototypical}, by considering $||\bm{r}_{H} - \bm{v}_k^H||_2^2$ as the cost of semantic mapping, this problem can be viewed as an optimal transmission problem.
In this setting, $q(c_{H}=k|\bm{r}_{H})$ represents the transmission or mapping scheme that needs to be solved.
In our implementation, we solve this equation by Sinkhorn-Knopp algorithm~\cite{cuturi2013sinkhorn}. The overall process of RQ with uniform semantic mapping is shown in Algorithm~\ref{alg:rq-usm}.

By optimizing the loss in Eqn.~\eqref{eq:rqloss}, we can obtain a trained encoder and multi-level codebooks.
During the construction of item indices, we first generate indices based on Eqn.~\eqref{eq:rqc}. 
After that, for each group of conflicting items, the codewords of these items at the last level will be redistributed uniformly based on Eqn.~\eqref{eq:rq-skh}. 
Such a two-stage process can also improve efficiency and reduce unnecessary random noise introduced by batching items~\cite{zhan2022learning}.

\subsection{Aligning Language and Collaborative Semantics in LLMs}
\label{sec:tuning}

After learning item indices, a straightforward approach is to integrate these index IDs into the LLM vocabulary, so that LLM can fulfill the recommendation task in a generative way that gradually omits the indices of items. 
However, these item indices are essentially OOV tokens for LLMs, and it is necessary to conduct the alignment between language and collaborative semantics. 
%Since the goal of our approach is to enable the recommendation capacity for LLMs, the key point lies in the alignment between language and collaborative semantics.   
%To integrate items among recommendation scenarios into the LLM, we first incorporate the additional tokens associated with item indices into the LLM vocabulary as OOV tokens.
For this purpose, we design a series of semantic alignment tasks to assign language and collaborative semantics for tuning LLMs, including the primary objective of sequential item prediction, explicit index-language alignment (identifying the corresponding item via their indices), and implicit recommendation-oriented alignment (enhancing comprehension of the language and collaborative semantics). 
As discussed below, these tuning tasks are very effective in enhancing the alignment between language models and collaborative semantics. 

%As our primary objective, sequential recommendation takes precedence as our initial focus. 
%It provides the LLM with fundamental recommendation abilities and aligns collaborative semantic information into item indices during the fine-tuning process.
%Secondly, to facilitate the integration of language semantics understandable by the LLM into item indices, we introduce a pair of explicit language semantic alignment tasks, which aids the LLM in precisely identifying items via their indices.
%Additionally, several implicit semantic alignment tasks are also introduced to further enhance the LLM‘s comprehension of the language and collaborative semantics associated with item indices.
%Finally, our aim is for the item index to serve as a universal identifier that seamlessly integrates both language and collaborative semantics.

\subsubsection{Sequential Item Prediction} \label{sec:seqrec} 
Since our approach is built in a LLM-based generative manner, we consider employing sequential item prediction as the major tuning objective. 
%The emergence of large language models such as ChatGPT and GPT-4 indicates that the generative paradigm is expected to become a universal approach for solving various tasks.  Specific to the sequential recommendation, there are three common paradigms: (1) Given a user's historical interactions, determine whether an item is suitable as the user's next item. (2) Sort a series of candidate items based on the user's historical interaction items. (3) Analyze user preferences based on his chronological behavior sequence and directly generate the next item for the user.
Specifically,  we construct personalized recommendation instructions based on the user's current historical interactions. Then, LLMs are prompted by the instructions and the interaction history, to predict the next item that the target user is likely to interact with. 
%In order to achieve full ranking on the entire item set, we focus on directly generating the index of the target item. In particular, we construct personalized recommendation instructions based on the user's current historical interactions. 
% For example ``\emph{Here are the user's historical interactions: \{history\}, try to recommend another item to the user. Note that the historical interactions are arranged in chronological order.}''. 
Here, the user's historical interactions are described and identified as an index sequence of interacted items arranged in chronological order. A sample instance is given as follows: 

\begin{center}
\begin{tcolorbox}
[colback=white,
colframe=black!70,
width=0.45\textwidth,
breakable,]
\begin{minipage}{\linewidth}
\begin{small}
\textbf{Instruction:} \\
Here are the user's historical interactions:\emph{\texttt{\begin{footnotesize}\textless{a\_124}\textgreater\textless{b\_192}\textgreater\textless{c\_41}\textgreater\textless{d\_17}\textgreater,...,\textless{a\_82}\textgreater\textless{b\_59}\textgreater\textless{c\_191}\textgreater\textless{d\_66}\textgreater\end{footnotesize}}},
try to recommend another item to the user. Note that the historical interactions are arranged in chronological order. \\
\textbf{Response:} \\
\begin{footnotesize}
\emph{\texttt{\textless{a\_112}\textgreater\textless{b\_32}\textgreater\textless{c\_5}\textgreater\textless{d\_175}\textgreater}}
\end{footnotesize}
\end{small}
\end{minipage}
\end{tcolorbox}
\end{center}

However, due to the large semantic gap, simply fine-tuning LLMs with the above target task, it is difficult to sufficiently integrate language and collaborative semantics in LLMs. 

%\textcolor{blue}{It seems that LLMs can be adapted for recommendation tasks through the generative paradigm as mentioned above, but just fine-tuning on the target recommendation task is actually shadow integration.  Since item indices, which are essentially OOV tokens, are not aligned with the semantic space of LLMs, the collaborative semantics implied by the recommender system are not integrated with the language semantics modeled by LLMs. Therefore, in addition to learning collaborative semantics from sequential recommendation, our next step is to further align item indices with the semantics of LLMs}

% \textcolor{blue}{The generative sequential recommendation task can be easily adapted to LLMs by designing instructions as described above or by directly stacking item indices~\cite{rajput2023recommender}.
% However, as OOV tokens, the index-related tokens necessitate learning the embedding representations from scratch. This presents a problem: all additional parameters are essentially based on learning collaborative filtering relationships from sequential recommendation data.
% In other words, item indices are not compatible with the LLM and have no connection with the original LLM semantics.
% This divergence from our initial direction aims to fully leverage the semantic comprehension and generation capabilities of LLMs. 
% Therefore, in addition to learning collaborative semantics from sequential recommendation, our next step is to further align item indices with language semantics.
% (Unclear, need rewriting)
%  }

\subsubsection{Explicit Index-Language Alignment} \label{sec:langali} 

Although our item indices are constructed based on titles and descriptions of items, they rely solely on shared prefix codewords to establish a weak correlation among items with similar language semantics. 
%Each index does not explicitly convey language semantic information about the item. \textcolor{blue}{Furthermore, the LLM is unable to identify an item by mapping its index to the corresponding title or description.} 
To further endow item indices with language semantics, we propose two explicit index-language alignment tasks for tuning LLMs. 
%a pair of tasks for explicit language semantic alignment.

On the one hand, the LLM should be capable of accurately identifying the item indices based on the associated title or description.
On the other hand, it is expected that LLM can naturally capture relevant item information from its indices.
 % As shown below, for the first perspective, we construct several instructions such as ``\emph{An item is called ``\{title\}'' and described as ``\{description\}'', can you tell me which item it is?}''. Instructs LLM to generate the corresponding item index according to the item's title/description or a combination of both.
% For the second perspective, the instructions we construct are similar to "\emph{Please tell me what item \{index\} is called, along with a brief description of it.}".
Considering the two aspects,  we first instruct the LLM to generate the corresponding item indices according to the item's title/description or a combination of both.
Then, we instruct the LLM to recover the item information based on its indices. 
We present two instruction samples to illustrate the two alignment tuning tasks in the following.   
%Together, these two perspectives jointly construct a pair of mutual conversion tasks between the item index and language information.

\begin{center}
\begin{tcolorbox}
[colback=white,
colframe=black!70,
width=0.45\textwidth,
breakable,]

\begin{minipage}{\linewidth}
\begin{small}
\textbf{Instruction:} \\
An item is called ``\emph{\texttt{\begin{footnotesize}Pokémon Moon - Nintendo 3DS\end{footnotesize}}}'' and described as ``\emph{\texttt{\begin{footnotesize}Pokémon Moon will launch in the ... become a Pokémon Champion!\end{footnotesize}}}'', can you tell me which item it is? \\
\textbf{Response:} \\
\emph{\texttt{\begin{footnotesize}\textless{a\_66}\textgreater\textless{b\_197}\textgreater\textless{c\_236}\textgreater\textless{d\_223}\textgreater\end{footnotesize}}}
\end{small}
\end{minipage}
\end{tcolorbox}
\end{center}

\begin{center}
\begin{tcolorbox}
[colback=white,
colframe=black!70,
width=0.45\textwidth,
breakable,]

\begin{minipage}{\linewidth}
\begin{small}
\textbf{Instruction:} \\
Please tell me what item \emph{\texttt{\begin{footnotesize}\textless{a\_66}\textgreater\textless{b\_197}\textgreater\textless{c\_236}\textgreater\textless{d\_223}\textgreater\end{footnotesize}}} is called, along with a brief description of it. \\
\textbf{Response:} \\
Item Title: \emph{\texttt{\begin{footnotesize}Pokémon Moon - Nintendo 3DS\end{footnotesize}}} \\
Item Description: \emph{\texttt{\begin{footnotesize}Pokémon Moon will launch in the ... become a Pokémon Champion!\end{footnotesize}}}
\end{small}
\end{minipage}
\end{tcolorbox}
\end{center}

Actually, such a mutual prediction method is essentially similar to what has been in cross-modal semantic alignment.
%This method of converting between index and language semantics is essentially equivalent to a specialized cross-modal semantic alignment. 
It can be an analogy with the mutual association or mapping between images and text~\cite{zhang2023speechgpt}, or the conversion between speech and text~\cite{jin2023unified}. 
By instruction tuning the LLM with these alignment instructions, item indices can be seamlessly integrated into the semantic space spanned by the LLM. 
%the newly introduced multi-modal tokens can seamlessly integrate into the LLM, serving as special words that the model can comprehend. Likewise, we align the explicit language semantics with the respective item index tokens by incorporating these mutual conversion tasks during the instruction tuning process.

% \subsubsection{Item Index as Universal Identifier}
\subsubsection{Implicit Recommendation-oriented Alignment} 
\label{sec:impli}
After being tuned with the above alignment tasks, LLMs can acquire basic knowledge of collaborative semantics. 
In this part, we further consider enhancing the model capacity via recommendation-oriented alignment tasks, so that LLM can better leverage both language and collaborative semantics to fulfill various recommendation tasks in a more accurate way. Specifically, we design the following three alignment tasks:  

%learning sequential recommendation tasks and mutual conversion tasks between the item index and explicit language information, item indices have acquired basic knowledge of collaborative and language semantics.

%Moreover, we take a step further by performing implicit semantic alignment around item indices and leveraging them for various recommendation-related scenarios. 
%This advancement aims to assist the LLM in implicitly unifying language and collaborative semantics with the item index while responding to user needs. First and foremost, it is desirable for the item index to be smoothly exchanged with the title and description in any given situation. 

%For this reason, we anticipate that sequential recommendation can still be effective even when the item index is substituted with either the item title or description.
%Secondly, users should be able to easily complete the query of target items according to their own intentions. That is, the LLM should have the ability to align the semantic information rich in the item index with user-specific needs. Thirdly, the user's historical behavior reflects his preferences, so we try to infer user preferences by analyzing the joint language and collaborative semantics in the item index sequence.
%Next, we will describe these three tasks in detail.

\paragraph{Asymmetric item prediction}
\label{sec:asy}
As discussed in Section~\ref{sec:seqrec}, for sequential item prediction,  both the interaction history (\emph{condition}) and the target item (\emph{target}) are formatted in the representation of item indices.
We call this tuning task \emph{symmetric} since both the condition and target for prediction are based on item indices. 
To further enhance the semantic alignment, we increase the prediction difficulty by changing the representations of condition and target, so as to derive different combinations of semantic representations for items.  Specially, we consider the following three representation methods: 
%Sequential recommendation primarily consists of two key elements: user interaction history and target item. Here, we try to combine different representations of these two elements to construct several asymmetric sequential recommendation tasks. 
(1) replacing the indices of target item with the item title, instructing the LLM to generate the item title directly based on the item index sequence;  
% Example instruction is as follows: ``\emph{Based on the user's historical interactions: \{history\}, try to predict the title of the item that the user may need next.}''.
(2) replacing the  indices of target item with the item description, instructing the LLM to generate the item features and attributes expected by the user;  
% Example instruction is as follows: ``\emph{Here is the item interaction history of the user: \{history\}, please tell me what features he expects from his next item.}''.
(3) representing the user interaction history as a text sequence of item titles instead of an index sequence, instructing the LLM to infer user preferences based on the title sequence.
% Example instruction is as follows: ``\emph{Given the title sequence of user historical interactive items: \{historical titles\}, can you recommend a suitable next item for the user?}''. Limited by the input length of LLM, we do not use description sequences to represent user interaction history.
The sample instructions for the  three scenarios are as follows:

\begin{center}
\begin{tcolorbox}
[colback=white,
colframe=black!70,
width=0.45\textwidth,
breakable,]

\begin{minipage}{\linewidth}
\begin{small}
\textbf{Instruction:} \\
Based on the user's historical interactions: \emph{\texttt{\begin{footnotesize}\textless{a\_38}\textgreater\textless{b\_94}\textgreater\textless{c\_198}\textgreater\textless{d\_59}\textgreater,...,\textless{a\_190}\textgreater\textless{b\_60}\textgreater\textless{c\_94}\textgreater\textless{d\_86}\textgreater\end{footnotesize}}}, try to predict the title of the item that the user may need next. \\
\textbf{Response:} \\
\emph{\texttt{\begin{footnotesize}NBA 2K16 - PlayStation 4\end{footnotesize}}}
\end{small}
\end{minipage}
\end{tcolorbox}
\end{center}

\begin{center}
\begin{tcolorbox}
[colback=white,
colframe=black!70,
width=0.45\textwidth,
breakable,]
\begin{minipage}{\linewidth}
\begin{small}
\textbf{Instruction:} \\
Here is the item interaction history of the user: \emph{\texttt{\begin{footnotesize}\textless{a\_38}\textgreater\textless{b\_94}\textgreater\textless{c\_198}\textgreater\textless{d\_59}\textgreater,...,\textless{a\_190}\textgreater\textless{b\_60}\textgreater\textless{c\_94}\textgreater\textless{d\_86}\textgreater\end{footnotesize}}}, please tell me what features he expects from his next item. \\
\textbf{Response:}\\
\emph{\texttt{\begin{footnotesize}The NBA 2K franchise is ... cover of your choice into the front of box.\end{footnotesize}}}
\end{small}
\end{minipage}
\tcbline
\begin{minipage}{\linewidth}
\begin{small}
\textbf{Instruction:} \\
Given the title sequence of user historical interactive items: \begin{footnotesize}``\emph{\texttt{The Biggest Loser Ultimate Workout - Xbox 360}}''\emph{\texttt{,...,}} ``\emph{\texttt{Lego: Marvel Super Heroes, XBOX 360}}''\end{footnotesize}, can you recommend a suitable next item for the user? \\
\textbf{Response:} \\
\emph{\texttt{\begin{footnotesize}\textless{a\_27}\textgreater\textless{b\_58}\textgreater\textless{c\_138}\textgreater\textless{d\_201}\textgreater \\
(Lego Star Wars - Xbox 360)\end{footnotesize}}}
\end{small}
\end{minipage}

\end{tcolorbox}
\end{center}

To make a comparison, the item prediction task in Section~\ref{sec:seqrec} 
%The previous generative sequential recommendation task in Section~\ref{sec:seqrec} 
involves mapping an index sequence to the target indices, while the tuning tasks in Section~\ref{sec:langali} explicitly align item indices with their corresponding language information. 
These asymmetric tasks are more difficult, which enforces LLMs to unify item indices, language semantics, and collaborative semantics for fulfilling the recommendation tasks.  As will be shown in the experiment part (Section~\ref{sec:ablation}), these tuning tasks are useful in adapting LLMs to recommender systems.  
%typically require the LLM to implicitly unify item index, language semantics, and collaborative semantics into a cohesive whole during the sequential recommendation process.
%Therefore, even though such asymmetric tasks may not hold practical significance in recommender systems, we still include them during training to implicitly assist in aligning item indices with language and collaborative semantics.

\paragraph{Item prediction based on user intention}
\label{sec:ir}
Drawing inspiration from~\cite{zhang2023recommendation}, a recommender system in real life should possess the ability to understand the actual intentions of users and provide high-quality recommendations accordingly. 
This leads to a task similar to item retrieval. %, which we also incorporate into our framework to further enhance LLM's understanding of language and collaborative semantics within item indices.
Referring to the approach in~\cite{zhang2023recommendation}, as reviews offer valuable evidence regarding users' personal tastes and motivations for making a specific interaction, we consider extracting intentions from the related reviews of the target item. To accomplish this, we utilize GPT-3.5 to process these reviews and extract user intentions.  
As for instructions, we mainly design two types of tasks:  
the former queries an item recommendation directly based on instant user intention, and the latter provides the user's interaction history for a personalized recommendation. 
% An example of instruction is ``\emph{Suppose you are a search engine, now a user searches that: ``\{query\}'', can you select an item to respond to the user's query?}''.
% An example of instruction is ``\emph{As a recommender system, you are assisting a user who has recently interacted with the following items: \{history\}. The user expresses a desire to obtain another item with the following characteristics: ``\{query\}''. Please recommend an item that meets these criteria.}''.
%It is worth noting that the retrieval target is the item index rather than the title as in~\cite{zhang2023recommendation}, \textcolor{blue}{which allows us to search for items that meet the user's needs from the entire item set.} 

\begin{center}
\begin{tcolorbox}
[colback=white,
colframe=black!70,
width=0.45\textwidth,
breakable,]

\begin{minipage}{\linewidth}
\begin{small}
\textbf{Instruction:} \\
Suppose you are a search engine, now a user searches that: ``\emph{\texttt{\begin{footnotesize}The game has an open world environment ... activities to complete\end{footnotesize}}}'', can you select an item to respond to the user's query? \\
\textbf{Response:} \\
\emph{\texttt{\begin{footnotesize}\textless{a\_104}\textgreater\textless{b\_4}\textgreater\textless{c\_47}\textgreater\textless{d\_182}\textgreater \\
(Grand Theft Auto Vice City Stories)\end{footnotesize}}}
\end{small}
\end{minipage}
\end{tcolorbox}
\end{center}

\begin{center}
\begin{tcolorbox}
[colback=white,
colframe=black!70,
width=0.45\textwidth,
breakable,]

\begin{minipage}{\linewidth}
\begin{small}
\textbf{Instruction:} \\
As a recommender system, you are assisting a user who has recently interacted with the following items: \emph{\texttt{\begin{footnotesize}\textless{a\_64}\textgreater\textless{b\_159}\textgreater\textless{c\_1}\textgreater\textless{d\_89}\textgreater,...,\textless{a\_119}\textgreater\textless{b\_98}\textgreater\textless{c\_162}\textgreater\textless{d\_155}\textgreater\end{footnotesize}}}. The user expresses a desire to obtain another item with the following characteristics: ``\emph{\texttt{\begin{footnotesize}The console offers 500GB of storage, ... 4K HDR gaming\end{footnotesize}}}''. Please recommend an item that meets these criteria. \\
\textbf{Response:} \\
\emph{\texttt{\begin{footnotesize}\textless{a\_227}\textgreater\textless{b\_206}\textgreater\textless{c\_156}\textgreater\textless{d\_156}\textgreater \\
(PlayStation 4 500GB Console)\end{footnotesize}}}
\end{small}
\end{minipage}
\end{tcolorbox}
\end{center}

%As for instructions, as shown above, we mainly design two types of tasks: item retrieval and personalized retrieval. The former queries an item directly based on user intention. 
% An example of instruction is ``\emph{Suppose you are a search engine, now a user searches that: ``\{query\}'', can you select an item to respond to the user's query?}''.
%The latter provides the user's interaction history for personalized retrieval. 
% An example of instruction is ``\emph{As a recommender system, you are assisting a user who has recently interacted with the following items: \{history\}. The user expresses a desire to obtain another item with the following characteristics: ``\{query\}''. Please recommend an item that meets these criteria.}''.
%It is worth noting that the retrieval target is the item index rather than the title as in~\cite{zhang2023recommendation}, \textcolor{blue}{which allows us to search for items that meet the user's needs from the entire item set.} 
%As we can see,  such a task is useful to align the implicit language and collaborative semantic information underlying the item index with user intentions rather than retrieval based on text similarity.

\paragraph{Personalized preference inference}
\label{sec:prefer}
Intuitively, a user's interaction history can implicitly reflect his or her personal preferences, but explicit preferences are generally absent from the dataset. Thus, we employ GPT-3.5  to infer the user's explicit preferences from items the user has interacted with in the past. %Then, the augmented data will be used to train our model. 
Unlike prior work~\cite{zhang2023recommendation},  we infer user preferences based on the index sequence of historical items rather than the title sequence.
% The instruction could be: "\emph{Utilizing the ordered list of the user's historical interaction items as a reference, please make an informed estimation of the user's preferences. The historical interactions are as follows: \{history\}.}" 
%\textcolor{blue}
This task requires the index sequence to act as an effective substitute for the title sequence, enabling the LLM to understand the joint language and collaborative semantics within the index sequence and accurately extract user preferences. 
The instruction could be:

\begin{center}
\begin{tcolorbox}
[colback=white,
colframe=black!70,
width=0.45\textwidth,
breakable,]
\begin{minipage}{\linewidth}
\begin{small}
\textbf{Instruction:} \\
Utilizing the ordered list of the user's historical interaction items as a reference, please make an informed estimation of the user's preferences. The historical interactions are as follows:
\emph{\texttt{\begin{footnotesize}\textless{a\_227}\textgreater\textless{b\_186}\textgreater\textless{c\_52}\textgreater\textless{d\_166}\textgreater,...,\textless{a\_120}\textgreater\textless{b\_208}\textgreater\textless{c\_146}\textgreater\textless{d\_153}\textgreater\end{footnotesize}}}. \\
\textbf{Response:} \\
\emph{\texttt{\begin{footnotesize}The user has recently been playing a game that ... with more data storage and/or higher data processing capabilities.\end{footnotesize}}}
\end{small}
\end{minipage}
\end{tcolorbox}
\end{center}

In this work, we mainly focus on the setting of sequential item prediction, \ie sequential recommendation, while our approach can be easily extended to other tuning tasks in recommender systems, \eg bundle prediction and explanation generation.   
Actually, our index mechanism can support various instruction tuning tasks as in standard language models~\cite{wang2023rethinking}, since these indices are endowed with both language and collaborative semantics, acting as common tokens for LLMs. 
%It is worth pondering that treating the item index as a universal identifier containing language and collaborative semantics can actually incorporate more implicit semantic alignment tasks into our framework. 
%%For instance, extending item retrieval tasks based on real-world user search logs, adding knowledge graph Q\&A tasks about items into the framework to further enrich the item background, explainable recommendations, etc.
%Here, due to some limitations in data acquisition and this paper primarily focuses on sequential recommendation, these tasks will only serve as auxiliary components. Therefore, the scope of the task will not be extended.

\subsection{Training and Inference} 
In this section, we discuss how to optimize our base LLM using the aforementioned tuning tasks and how to utilize it to fulfill the recommendation task over the entire item set. 

%we begin by introducing the LLM that we utilize as our backbone model. 
%Subsequently, we discuss how to optimize our base LLM using the aforementioned tasks and how to utilize it to address the recommendation task across the entire item set.

\subsubsection{Training} 
In this paper, we strive to leverage the semantic understanding and generation capabilities of LLMs to facilitate sequential recommendation. To this end, we employ LLaMA~\cite{touvron2023llama} as our backbone model and then optimize it via instruction tuning.
The tuning tasks mentioned above can be conveniently formatted as conditional language generation tasks in a sequence-to-sequence manner. We optimize the negative log-likelihood of the generation target as follows:
\begin{align}
    \mathcal{L} = - \sum_{\langle I,Y \rangle \in \bm{B}} \sum_{j=1}^{|Y|}\operatorname{log}P(Y_j|I,Y_{<j}),
    \label{eq:llmloss}
\end{align}
where $\langle I,Y \rangle$ represents a pair of instruction and target response in the batch data, $Y_j$ is the $j$-th token of $Y$ and $Y_{<j}$ denotes the tokens before $Y_j$. 
For each task, we designed multiple instruction templates to enhance the instruction diversity. 
However, during a training epoch, each data is only combined with one sampled instruction template, which is different from those in prior approaches~\cite{geng2022recommendation,hua2023index,zhang2023recommendation}. 
% This decision 
This strategy is based on our observation that LLaMA, as an LLM with over 7B parameters, achieves better results by examining specific data only a few times~\cite{lee2022deduplicating}. In contrast, repeating data may lead to overfitting.
% multiple instances of the same data within an epoch will lead to significant overfitting.

% \textcolor{blue}{Prior  approaches~\cite{geng2022recommendation,hua2023index,zhang2023recommendation} often sampled several instruction templates for each piece of data in one epoch. 
% This decision is based on our observation that LLaMA, as a LLM with a larger parameter quantity, achieves better results by examining specific data only a few times. However, multiple instances of the same data within an epoch will lead to significant overfitting.
% (need rewriting)
% }

\subsubsection{Inference} 
Our objective is to generate the top $n$ items from the entire item set which most match the preference of a given user during inference. To accomplish this, the decoder module performs a beam search across the index tokens.
Here, we use the index structure built in Section~\ref{sec:index} for item decoding. 
% \textcolor{blue}{Additionally, at each step of the item decoding, we restrict the generation range to the corresponding level of index IDs. This restriction helps minimize the occurrence of invalid items.}
Additionally, when calculating logits, the probabilities of tokens that may result in illegal item indices will be assigned as 0 to ensure generation quality.

Given an input sequence, the inference time is mainly consumed in the multi-layer self-attention calculation.
The time complexity of a forward process in the vanilla Transformer is $\mathcal{O}(N^2dL)$, where $L$ is the number of model layers, $N$ is the sequence length, and $d$ is the dimension of hidden states. Overall, in order to autoregressively generate complete target item indices, the time consumption is $\mathcal{O}(HN^2dL)$,  where $H$ is the number of index levels (usually a small value like $H=4$). 
But in fact, the attention key and value tensors of each layer can be cached for subsequent decoding, called \emph{KV Cache}~\cite{pope2023efficiently}. After applying \emph{KV Cache}, the time complexity can be optimized to $\mathcal{O}(N^2dL + HNdL)$.
In addition to the inference speed, the memory efficiency can also be improved through various technologies such as model quantization~\cite{gholami2021survey} and PagedAttention~\cite{kwon2023efficient}.
% Assuming that the beam size is $N$ and the item index length is $L$, the time complexity of the above inference process is $\mathcal{O}(NL)$. But in fact, the values of $N$ and $L$ in our scenarios are often very small (\eg, $N=20$ and $L=4$), and the generation efficiency can also be improved through various technologies such as model parameter compression~\cite{gholami2021survey} and PagedAttention~\cite{kwon2023efficient}. Below we briefly show the inference efficiency of the model under a NVIDIA A100 SXM4 40GB GPU in Table~\ref{}.

% \begin{table}
% \centering
% \caption{}
% \label{tab:infer}
% \setlength{\tabcolsep}{1.5mm}{
% \begin{tabular}{lllcccc}
% \toprule

% \bottomrule
% \end{tabular}}
% \end{table}

% Additionally, we impose constraints that the model should generate corresponding \textcolor{blue}{level index tokens at each step} to minimize the occurrence of illegal items.

\subsection{Discussion} 
\label{sec:discussion}

In this part, we compare the proposed LC-Rec with existing language model based methods for recommendation to highlight the contributions of our approach.

\textbf{Text-based methods} such as TALLRec~\cite{bao2023tallrec} and InstructRec~\cite{zhang2023recommendation} typically represent user historical behavior as a sequence of item titles, thereby formatting the sequential recommendation task into a natural language question or instruction, which can be easily adapted to the LLM.
However, these methods are not suitable for the full ranking setting, since they have difficulty in understanding and generating the item information over the entire item set. %across the entire item set due to generation difficulties caused by the length and diversity of item titles.
They either consider a discrimination question that can be answered with ``\emph{Yes/No}''~\cite{bao2023tallrec} or perform reranking based on a small number of candidate items~\cite{zhang2023recommendation}.
Furthermore, this approach mainly relies on language semantics to tackle the recommendation tasks, which neglects the collaborative semantic information in recommender systems. 
%straightforward approach to natural language organization overlooks the potential benefits of incorporating collaborative semantic information in recommender systems.
%Different from those methods, our approach applies to full ranking on the entire item set. Besides, the alignment of language and collaborative semantics in the LLM tuning facilitates the effective utilization of the semantic comprehension and generation capabilities of LLM in sequential recommendations.

\begin{table}
\centering
\caption{Comparison of our method with several related studies. ``FR'' denotes full ranking across the entire item set. ``LS'' denotes language semantics. ``CS'' denotes collaborative semantics. ``ILC'' denotes the integration of language and collaborative semantics in LLMs.}
\label{tab:comp}
\setlength{\tabcolsep}{1.5mm}{
\begin{tabular}{lllcccc}
\toprule
Methods &Scale & Backbone & FR &  LS &  CS  &  ILC\\
\midrule
TIGER~\cite{rajput2023recommender} & N/A & N/A & \textcolor[RGB]{0,139,0}{\CheckmarkBold} & \textcolor{red}{\XSolidBrush} & \textcolor[RGB]{0,139,0}{\CheckmarkBold} & \textcolor{red}{\XSolidBrush}\\
P5~\cite{geng2022recommendation,hua2023index} &220M  & T5 & \textcolor[RGB]{0,139,0}{\CheckmarkBold} & \textcolor{red}{\XSolidBrush} & \textcolor[RGB]{0,139,0}{\CheckmarkBold} & \textcolor{red}{\XSolidBrush}\\
% CLLM4Rec~\cite{zhu2023collaborative} & 1.5B & GPT-2 & \textcolor[RGB]{0,139,0}{\CheckmarkBold} & \textcolor[RGB]{0,139,0}{\CheckmarkBold} & \textcolor[RGB]{0,139,0}{\CheckmarkBold} & \textcolor{red}{\XSolidBrush}\\
InstructRec~\cite{zhang2023recommendation} & 3B & Flan-T5 & \textcolor{red}{\XSolidBrush} & \textcolor[RGB]{0,139,0}{\CheckmarkBold} & \textcolor{red}{\XSolidBrush} & \textcolor{red}{\XSolidBrush}\\
TALLRec~\cite{bao2023tallrec} & 7B & LLaMA & \textcolor{red}{\XSolidBrush} & \textcolor[RGB]{0,139,0}{\CheckmarkBold} & \textcolor{red}{\XSolidBrush} & \textcolor{red}{\XSolidBrush}\\
LC-Rec & 7B & LLaMA & \textcolor[RGB]{0,139,0}{\CheckmarkBold} &\textcolor[RGB]{0,139,0}{\CheckmarkBold} & \textcolor[RGB]{0,139,0}{\CheckmarkBold} & \textcolor[RGB]{0,139,0}{\CheckmarkBold} \\
\bottomrule
\end{tabular}}
\end{table}

\textbf{Index-based methods} such as TIGER~\cite{rajput2023recommender} and P5~\cite{geng2022recommendation,hua2023index} (specifically, we focus on the sequential recommendation task of P5) directly convert the traditional item ID-based sequential recommendation into a generative paradigm.
TIGER is not based on language models. Instead, it trains an encoder-decoder Transformer model from scratch to predict the next item given an input item sequence, where each item is identified by multiple discrete IDs.
P5 adapts recommendation tasks into the text-to-text format to enable unified modeling. 
However, within this framework, sequential recommendation is still organized as a mapping process from item ID sequence to target item ID, which only establishes collaborative semantics between item IDs and is independent of language semantics in LLMs.
We are also aware of several concurrent studies~\cite{lin2023multifacet,zhang2023collm,zhu2023collaborative} that aim to adapt LLMs for recommender systems. They mainly consider enhancing the semantics of items from different aspects, including setting multi-type identifiers (\eg ID, title, and attributes)~\cite{lin2023multifacet}, incorporating external collaborative representations~\cite{zhang2023collm}, and learning dual user/item embeddings~\cite{zhu2023collaborative}.

As a comparison, our work focuses on the integration of language and collaborative semantics for enhancing the recommendation capacity of LLMs. 
Specifically, we adopt a new item indexing mechanism that ensures index uniqueness and effectively reduces the vocabulary size. Moreover, we further design various alignment tasks for enhancing the semantic integration. Based on these improvements, our approach can effectively integrate collaborative semantics into LLMs, and further leverage the enhanced capacity of LLMs for fulfilling the recommendation tasks. 
The comparison of our method with several related studies is shown in Table~\ref{tab:comp}.

\section{Experiment}
In this section, we first set up the experiments and then present the results as well as analysises of our proposed approach.

\subsection{Experiment Setup}

\begin{table}[]
\centering
\caption{Statistics of the preprocessed datasets. ``\textbf{Avg.} \emph{len}'' represents the average length of item sequences.}
\label{tab:data_statistics}
\setlength{\tabcolsep}{1.7mm}{
\begin{tabular}{lrrrrr}
\toprule
 \textbf{Datasets} & \textbf{\#Users} & \textbf{\#Items}  & \textbf{\#Interactions} & \textbf{Sparsity} & \textbf{Avg.} \emph{len}\\
\midrule
Instruments & 24,773 & 9,923  & 206,153  & 99.92\%  & 8.32\\
Arts & 45,142 & 20,957 & 390,832   & 99.96\%  & 8.66\\
Games & 50,547 & 16,860 & 452,989 & 99.95\%  & 8.96\\ 
\bottomrule
\end{tabular}}
\end{table}

\subsubsection{Dataset}

We evaluated the proposed approach on three subsets of Amazon review data~\cite{ni2019justifying}, including ``\emph{Musical Instruments}'', ``\emph{Arts, Crafts and Sewing}'', and ``\emph{Video Games}''. All three datasets contain user review data from May 1996 to October 2018. Each item in the dataset is associated with a title and a description. Following previous work~\cite{hou2022towards}, we first filter out unpopular users and items with less than five interactions. Then, we create user behavior sequences based on the chronological order. The maximum item sequence length is uniformly set to 20 to meet all baseline requirements.
The statistics of our preprocessed datasets are shown in Table~\ref{tab:data_statistics}.

\begin{table*}[]
\centering
\caption{Performance comparison of different methods on the three datasets. The best and second-best performances are indicated in bold and underlined font, respectively. 
The performance for our LC-Rec is average results from multiple instruction templates.}
\label{tab:res}
% \centering
\renewcommand\arraystretch{1.3}
\setlength{\tabcolsep}{0.6mm}{
\begin{tabular}{l|l|p{1.1cm}<{\centering}p{1.1cm}<{\centering}p{1.1cm}<{\centering}p{1.2cm}<{\centering}p{1.1cm}<{\centering}p{1.28cm}<{\centering}|p{1.1cm}<{\centering}p{1.1cm}<{\centering}|p{1.1cm}<{\centering}p{1.1cm}<{\centering}|p{1.1cm}<{\centering}p{1.1cm}<{\centering}}
\hline
Dataset  & Metrics   & Caser  & HGN    & GRU4Rec & BERT4Rec & SASRec & FMLP-Rec & FDSA & S$^3$-Rec & P5-CID  &TIGER & LC-Rec &Improv. \\ 
\hline
\multirow{5}{*}{Instruments} & HR@1  & 0.0149 & 0.0523 & 0.0571  & 0.0435   & 0.0503 & 0.0480   & 0.0520& 0.0367& 0.0587
& \underline{0.0608}
& \textbf{0.0706}  & +16.12\%
\\
                             & HR@5  & 0.0543 & 0.0813 & 0.0821  & 0.0671   & 0.0751 & 0.0786   & 0.0834& \underline{0.0863} & 0.0827
&\underline{0.0863}
& \textbf{0.1002} & +16.11\%
\\
                             & HR@10 & 0.0710 & 0.1048 & 0.1031  & 0.0822   & 0.0947 & 0.0988   & 0.1046& \underline{0.1136}& 0.1016
&0.1064
& \textbf{0.1220}  & +7.39\%
\\
                             & NDCG@5    & 0.0355 & 0.0668 & 0.0698  & 0.0560   & 0.0627 & 0.0638   & 0.0681& 0.0626& 0.0708
& \underline{0.0738}
& \textbf{0.0856}  & +15.99\%
\\
                             & NDCG@10   & 0.0409 & 0.0744 & 0.0765  & 0.0608   & 0.0690 & 0.0704   & 0.0750& 0.0714& 0.0768
& \underline{0.0803}
& \textbf{0.0926}  & +15.32\%
\\
\hline
\multirow{5}{*}{Arts}        & HR@1  & 0.0138 & 0.0300 & 0.0421  & 0.0337   & 0.0225 & 0.0310   & 0.0451& 0.0245& \underline{0.0485}
&0.0465
& \textbf{0.0634} & +30.72\%
\\
                             & HR@5  & 0.0379 & 0.0622 & 0.0749  & 0.0559   & 0.0757 & 0.0757   & 0.0734& 0.0767& 0.0724
& \underline{0.0788}
& \textbf{0.1011}  & +28.30\%
\\
                             & HR@10 & 0.0541 & 0.0875 & 0.0964  & 0.0713   & 0.1016 & 0.1046   & 0.0933& \underline{0.1051} & 0.0902
&0.1012
& \textbf{0.1266}  & +20.46\%
\\
                             & NDCG@5    & 0.0262 & 0.0462 & 0.0590  & 0.0451   & 0.0508 & 0.0541   & 0.0595& 0.0521& 0.0607
& \underline{0.0631}
& \textbf{0.0828}  & +31.22\%
\\
                             & NDCG@10   & 0.0313 & 0.0544 & 0.0659  & 0.0500   & 0.0592 & 0.0634   & 0.0660& 0.0612& 0.0664
& \underline{0.0703}
& \textbf{0.0906}  & +28.88\%
\\
\hline
\multirow{5}{*}{Games}       & HR@1  & 0.0085 & 0.0154 & 0.0176  & 0.0136
& 0.0145 & 0.0152   & 0.0161& 0.0119& 0.0177
& \underline{0.0188}
& \textbf{0.0317}  & +68.62\%
\\
                             & HR@5  & 0.0367 & 0.0517 & 0.0586  & 0.0482
& 0.0581 & 0.0571   & \underline{0.0644} & 0.0606& 0.0506
&0.0599
& \textbf{0.0800}  & +24.22\%
\\
                             & HR@10 & 0.0617 & 0.0856 & 0.0964  & 0.0763
& 0.0940 & 0.0930   & \underline{0.1041} & 0.1002& 0.0803
&0.0939
& \textbf{0.1174}  & +12.78\%
\\
                             & NDCG@5    & 0.0227 & 0.0333 & 0.0381  & 0.0311
& 0.0365 & 0.0361   & \underline{0.0404} & 0.0364& 0.0342
&0.0392
& \textbf{0.0560}  & +38.61\%
\\
                             & NDCG@10   & 0.0307 & 0.0442 & 0.0502  & 0.0401 
& 0.0481 & 0.0476   & \underline{0.0531} & 0.0491 & 0.0437 &0.0501 
& \textbf{0.0681}  & +28.25\%
\\
\hline
\end{tabular}}
\end{table*}

\subsubsection{Baseline Models}

We adopt the following representative sequential recommendation models as baselines for comparison with our LC-Rec:
\begin{itemize}
\item {\textbf{Caser}}~\cite{tang2018personalized} is a CNN-based method that models user behaviors by applying horizontal and vertical convolutional filters.
\item {\textbf{HGN}}~\cite{ma2019hierarchical} utilizes hierarchical gating networks to capture both long-term and short-term user interests from historical behaviors.
\item {\textbf{GRU4Rec}}~\cite{hidasi2015session} is an RNN-based sequential recommendation model that utilizes GRU to encode the item sequence.
\item {\textbf{BERT4Rec}}~\cite{sun2019bert4rec} adopts a bidirectional Transformer model and combines it with a mask prediction task for the modeling of item sequences.
\item {\textbf{SASRec}}~\cite{kang2018self} exploits a unidirectional Transformer-based neural network to model the item sequences and predict the next item.
\item {\textbf{FMLP-Rec}}~\cite{zhou2022filter} proposes an all-MLP model with learnable filters, which ensures efficiency and reduces noise signals.
\item {\textbf{FDSA}}~\cite{zhang2019feature} focuses on the transformation patterns between item features, modeling both item-level and feature-level sequences separately through self-attention networks.
\item {\textbf{S$^3$-Rec}}~\cite{zhou2020s3} utilizes mutual information maximization to pre-train a self-supervised sequential recommendation model, learning the correlation between items and attributes.
\item {\textbf{P5-CID}}~\cite{geng2022recommendation,hua2023index} organizes multiple recommendation tasks in a text-to-text format and models different tasks uniformly using the T5 model. Subsequently, the author team explores the construction of item indices for sequential recommendation, including sequential indexing and collaborative indexing. Here, we employ P5 with collaborative indexing as the baseline and implement it according to the code\footnote{https://github.com/Wenyueh/LLM-RecSys-ID/} provided by the authors.
\item {\textbf{TIGER}}~\cite{rajput2023recommender} adopts the generative retrieval paradigm for sequential recommendation and introduces a semantic ID to uniquely identify items. Due to the official code not being released by the authors, here we implement it ourselves by Transformers\footnote{https://github.com/huggingface/transformers/ \label{fn:hf}} following the implementation details provided in the paper.
\end{itemize}

\subsubsection{Evaluation Settings}
To evaluate the performance of sequential recommendation, we adopt two widely used metrics, top-$K$ Hit Ratio (HR) and top-$K$ Normalized Discounted Cumulative Gain (NDCG). In this paper, we set $K$ as 1, 5, and 10. Following previous works~\cite{kang2018self,zhou2020s3,zhou2022filter}, we employ the \emph{leave-one-out} strategy for evaluation. Concretely, for each user behavior sequence, the most recent item is used as the test data, the second most recent item is used as the validation data, and the remaining interaction records are used for training. We perform full ranking evaluation over the entire item set instead of sample-based evaluation. For the generative methods based on beam search, the beam size is uniformly set to 20.

\subsubsection{Implementation Details}
To construct item indices, we utilize LLaMA to encode the title and description of the item as its embedding and use mean pooling to aggregate multiple token representations. 
The level of item indices is set to 4, with each level consisting of 256 codebook vectors, and each vector has a dimension of 32.
Both the encoder and decoder of RQ-VAE are implemented as Multi-Layer Perceptrons (MLPs) with ReLU activation functions. The model is optimized using the AdamW optimizer, employing a learning rate of 0.001 and a batch size of 1024.

For LLM fine-tuning, we implemented LC-Rec based on LLaMA through Transformers\footref{fn:hf} and accelerated training by DeepSpeed\footnote{https://github.com/microsoft/DeepSpeed/}. 
All tokens related to item indices are appended to the tokenizer as out-of-vocabulary (OOV) tokens. 
We employ the AdamW optimizer for model optimization, setting the learning rate to 5e-5 and weight decay to 0.01. 
During the fine-tuning, a cosine scheduler with warmup is utilized to adjust the learning rate. 
With the application of data parallelism and gradient accumulation, the overall batch size amounts to 128. We conduct training for 4 epochs on each dataset. To prevent overfitting, we ensure that during each epoch, a single data is combined with one sampled instruction template and appears only once.

\begin{table*}
\centering
\caption{Ablation study of various semantic alignment tasks in LC-Rec. We show the results on Arts and Games dataset.}
\label{tab:ablation}
\renewcommand\arraystretch{1.3}
\setlength{\tabcolsep}{1.8mm}{
\begin{tabular}{l|p{1.2cm}<{\centering}p{1.2cm}<{\centering}p{1.2cm}<{\centering}p{1.2cm}<{\centering}p{1.2cm}<{\centering}|p{1.2cm}<{\centering}p{1.2cm}<{\centering}p{1.2cm}<{\centering}p{1.2cm}<{\centering}p{1.2cm}<{\centering}}  
\hline
\multirow{2}{*}{Methods} & \multicolumn{5}{c}{Arts}  & \multicolumn{5}{c}{Games}  \\ 
\cline{2-6} \cline{7-11} 
& HR@1 & HR@5 & HR@10 & NDCG@5 & NDCG@10 & HR@1 & HR@5 & HR@10 & NDCG@5 & NDCG@10 \\
\hline
SEQ & 0.0561 & 0.0909 & 0.1133 & 0.0740 & 0.0812 & 0.0243 & 0.0626 & 0.0930 & 0.0437 & 0.0535 \\
\ + MUT & 0.0593 & 0.0926 & 0.1141 & 0.0765 & 0.0832 & 0.0275 & 0.0703 & 0.1038 & 0.0491 & 0.0598 \\
\ + ASY & 0.0602 & 0.0945 & 0.1172 & 0.0776 & 0.0848 & 0.0281 & 0.0725 & 0.1073 & 0.0506 & 0.0615 \\
\ + ITE & \textbf{0.0638} & 0.0996 & 0.1232 & 0.0813 & 0.0889 & 0.0294 & 0.0770 & 0.1125 & 0.0534 & 0.0648 \\
\ + PER & 0.0634 & \textbf{0.1011} & \textbf{0.1266} & \textbf{0.0828} & \textbf{0.0906} & \textbf{0.0317} & \textbf{0.0800} & \textbf{0.1174} & \textbf{0.0560} & \textbf{0.0681} \\
\hline
\end{tabular}}
\end{table*}

\subsection{Overall Performance}
We compare the proposed approach with the different baseline models on three datasets, and the overall results are shown in Table~\ref{tab:res}. Based on these results, we can find:

For the baseline methods, the sequential recommendation methods that incorporate item content information (\ie FDSA and S$^3$-Rec) perform better than traditional sequential recommendation methods that solely rely on ID and collaborative relationships (\ie Caser, HGN, GRU4Rec, BERT4Re, SASRec, FMLP-Rec) on several datasets. 
This indicates that item content information introduced as additional information can effectively improve recommendation performance.
% This is because additional item-related information is used as auxiliary features to improve performance. 
As for P5-CID and TIGER, they demonstrate competitive performance across the first two datasets, particularly excelling in HR@1 and the metrics related to item ranking (\ie NDCG).
% 
% However, their advances in the HR@10 metric are not significantly pronounced, which we suspect could be influenced by the beam size used in beam search inference. 
In terms of the Games dataset, they have an improvement compared to the ID-only model, but no significant improvement compared to the methods that already include auxiliary content information. One possible reason for this is the difference in the effects of content information and the difficulty of modeling it in different data and scenarios.

% In terms of the third Games dataset, neither of them showed a notable improvement compared to the methods that incorporate item auxiliary information. This further emphasizes the importance of item-related background information in certain recommendation scenarios.

Our proposed LC-Rec consistently maintains the best performance on three datasets and shows significant improvements compared to the baseline methods. This superior performance can be attributed to two factors: (1) The item indexing mechanism via vector quantization combined with uniform semantic mapping, which captures similarities between items and ensures a semantically lossless generation process at the last index level. (2) The effective integration of collaborative semantics into LLMs, which results in a seamless fusion of language semantics and collaborative semantics. By employing these strategies, our approach is able to leverage the powerful modeling capabilities of LLMs, thereby achieving significant improvements in the recommendation task.

% effectively aligning language and collaborative semantics with item indices through various semantic alignment tasks, elegantly integrating items among the recommendation scenario into LLM. This integration fully leverages the semantic comprehension and generation capabilities of LLM, resulting in better performance.

\subsection{Ablation Study}
\label{sec:ablation}
% sequential item prediction, explicit index-language alignment (identifying the corresponding item via their indices), and implicit recommendation-oriented alignment (enhancing comprehension of the language and collaborative semantics)
\paragraph{Various semantic alignment tasks}
Our proposed LC-Rec consists of various semantic alignment tasks, including 
(1) \underline{SEQ}: the sequential item prediction task introduced in Section~\ref{sec:seqrec} as our primary objective, 
(2) \underline{MUT}: the mutual prediction task for explicit index-language alignment in Section~\ref{sec:langali}, 
(3) \underline{ASY}: the asymmetric item prediction task in Section~\ref{sec:asy}, 
(4) \underline{ITE}: the item prediction based on user intention in Section~\ref{sec:ir}, 
(5) \underline{PER}: the personalized preference inference task in Section~\ref{sec:prefer}.
The latter three tasks all belong to the implicit recommendation-oriented alignment introduced in Section~\ref{sec:impli}.
To validate the effectiveness of each component, we conduct the ablation study on Arts and Games dataset to analyze the contribution of each part. 

The results, as shown in Table~\ref{tab:ablation}, indicate that the gradual incorporation of multiple semantic alignment tasks into the sequential recommendation, which involves only collaborative semantics, can significantly improve performance.
% From the results in Table~\ref{tab:ablation}, we can observe that gradually incorporating various semantic alignment tasks to the sequential recommendation involving only collaborative semantics can bring great performance improvement.
All these instruction tuning tasks in LC-Rec are shown beneficial for enhancing sequential recommendation, and there is potential for further improvements by adding more semantic alignment tasks.

\paragraph{Other item indexing methods}
In addition to the semantic alignment tasks, we also examine the proposed item indexing method, by comparing it to another three indexing methods. 
(1)~\underline{Vanilla ID} is the same as the traditional recommendation model, using a single and unique ID for each item.
(2)~\underline{Random Indices} uses multi-level indexing, but the indices at each level are derived from random sampling and are not semantically related.
(3)~\underline{LC-Rec w/o USM} removes the uniform semantic mapping in our indexing method and assigns a distinct supplementary index ID to each conflicting item.

% One approach still exploits a multi-level item index, but each level of index is obtained through random sampling independent of semantics (called \underline{Random Indices}).
% The other approach resembles traditional recommendation models, utilizing a single vanilla ID to represent each item (called \underline{Vanilla ID}).

\begin{figure}[]
\centering
\includegraphics[width=0.46\textwidth]{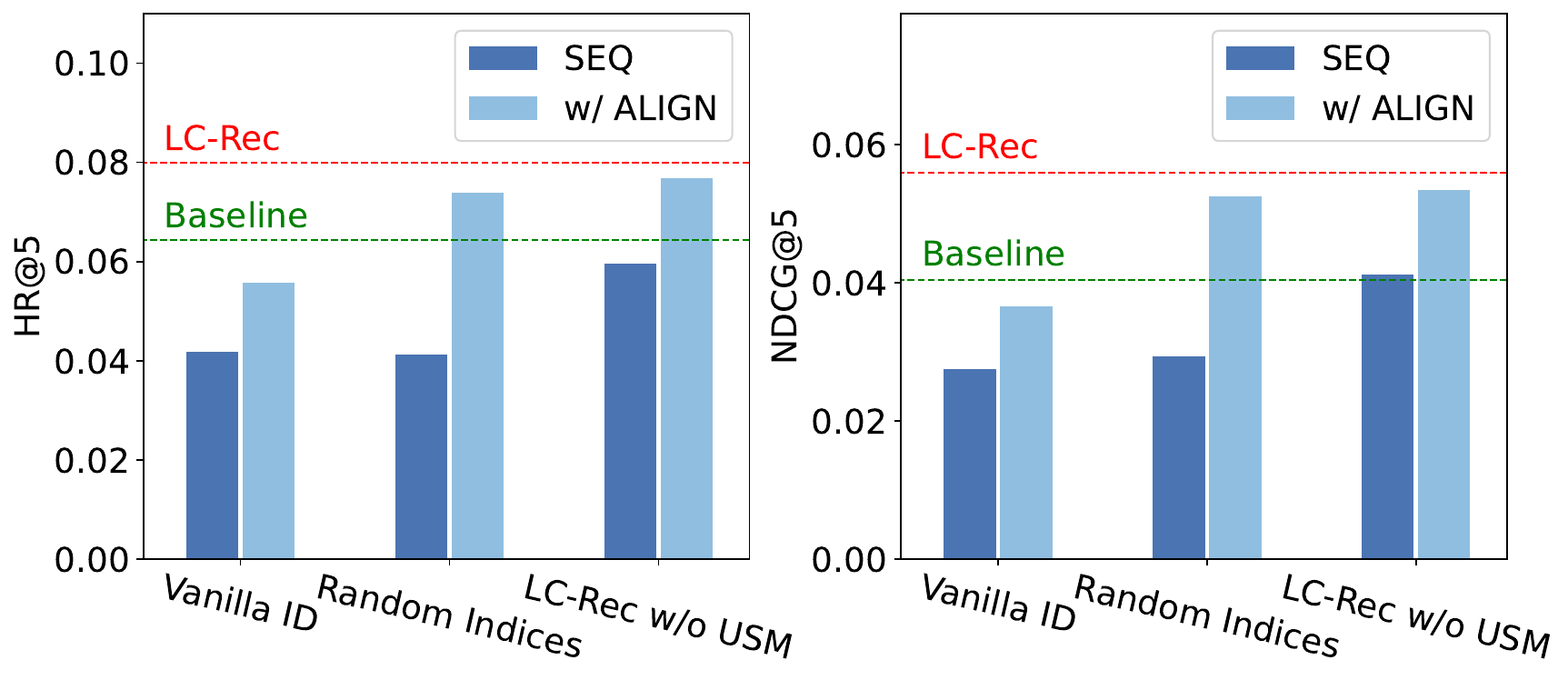}
\caption{The performance of our framework on three indexing methods, we report HR@5 and NDCG@5 on Games dataset. ``SEQ'' denotes fine-tuning only with the sequential item prediction task. ``w/ ALIGN'' denotes combining with our semantic alignment tasks.}
\label{fig:index}
\end{figure}

As shown in Figure~\ref{fig:index}, our approach (red dotted line, LC-Rec) outperforms all three base indexing methods, indicating the effectiveness of the proposed item indexing method (Section~\ref{sec:index}).
% In addition, 
% it is worth noting that our experimental results also indicate the ability of our proposed framework to adapt to multiple-item indexing mechanisms and reduce their sensitivity to the final performance.
In addition, if we apply the proposed semantic alignment tasks (``w/ ALIGN'' in Figure~\ref{fig:index}) to these three base indexing methods, their performance can be boosted by a large margin, especially for methods also based on multi-level indexing (\eg \underline{Random Indices} and \underline{LC-Rec w/o USM}), outperforming all baseline methods. The results also demonstrate that the proposed alignment tasks can improve recommendation performance in an indexing-agnostic way.
% when fine-tuning only with the sequential item prediction task, the three base indexing methods fail to achieve comparable results to the baselines. However, when combined with our semantic alignment tasks, their performance is substantially improved, especially for the \underline{Random Indices} method, which outperforms all baseline models.}
% 
% From the results in Figure~\ref{fig:index}, it can be seen that our method exhibits great advantages compared to these two suboptimal indexing methods. 
% Furthermore, it is worth noting that the experimental results also demonstrate that our framework is able to mitigate the sensitivity to the item indexing mechanism.
% If fine-tuning only with the target sequential item prediction task, these two suboptimal indexing methods cannot even achieve results close to the baselines. 
% However, when combined with our semantic alignment tasks, significant performance improvements are achieved, especially the random multi-level index, which surpasses all baseline models.

\subsection{Further Analysis}

\paragraph{Item prediction based on user intention}

We further evaluate the ability of LC-Rec to understand the semantics contained in the item index. The evaluation is performed through a user intention-based item prediction task on Games dataset, as described in Section~\ref{sec:ir}. 
Following the widely used setups in sequential recommendation task, the most recent record in each user behavior sequence is used for testing. User intentions are used as the query and are generated by GPT-3.5 based on review data. We employ DSSM~\cite{huang2013learning}, a widely validated retrieval model, as our baseline. It adopts a two-tower architecture to search for relevant items based on textual similarity between a given user query and item titles. In our implementation, BERT~\cite{kenton2019bert} is used to encode queries and item titles.

% In this part, we further evaluate LC-Rec's ability to understand the semantics within item indices through the task introduced in Section~\ref{sec:ir} (\ie Item prediction based on user intention).
% Following the same principle as in sequential recommendation, the most recent record of each user behavior sequence is used for testing. 
% As described in Section ~\ref{sec:ir}, user intentions as queries are generated by GPT-3.5 based on review data.
% We use DSSM~\cite{huang2013learning} as our baseline model, which is a widely verified retrieval model. It adopts a two-tower architecture to search for the relevant item based on text similarity between the given user query and item titles. In our implementation, BERT~\cite{kenton2019bert} is used to encode queries and item titles.

\begin{figure}[]
\centering
\includegraphics[width=0.47\textwidth]{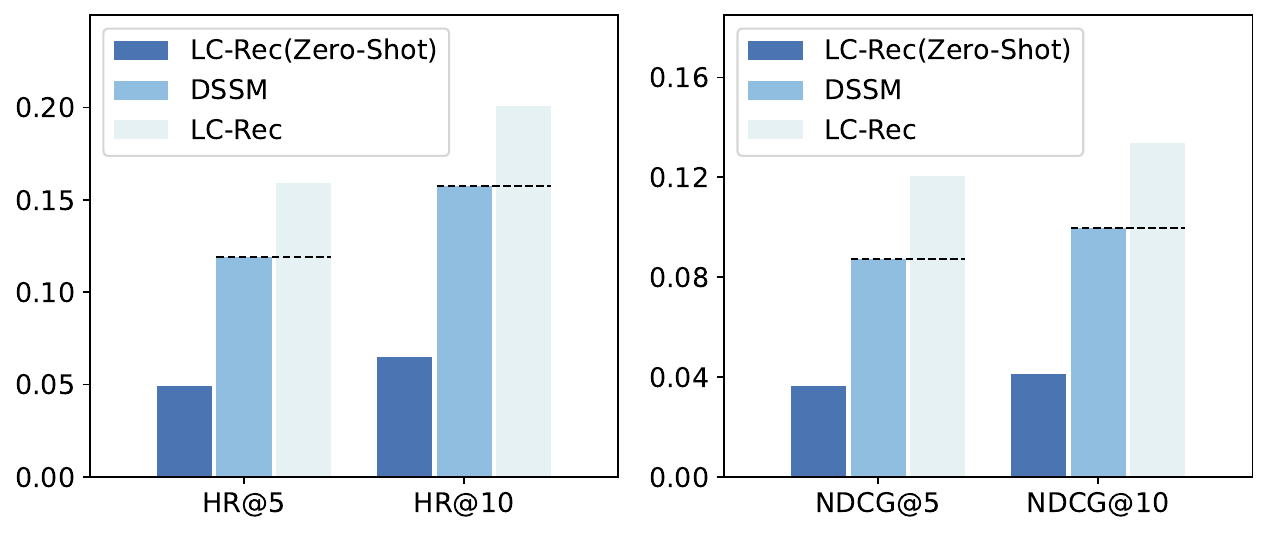}
\caption{Performance of item prediction based on user intention.}
\label{fig:ir}
\end{figure}

As shown in Figure~\ref{fig:ir}, our approach exhibits a significant performance improvement compared to the baseline model. 
This improvement can be attributed to the integration of language and collaborative semantics in the LLM through item indices.
Additionally, ``LC-Rec (Zero-Shot)'' represents the LC-Rec variant that is not trained in the item prediction task regarding user intention. Interestingly, we can observe that basic language and collaborative semantic alignment can still link item indices to user intentions to some extent, even without prior training on the target task.
% It can be observed that even without prior exposure to the target task, basic language and collaborative semantic alignment can still match item indices to some extent with user intentions.

\begin{figure}[]
\centering
\subfigure[Fine-tuning only with the target sequential item prediction task.]{\includegraphics[width=0.485\linewidth]{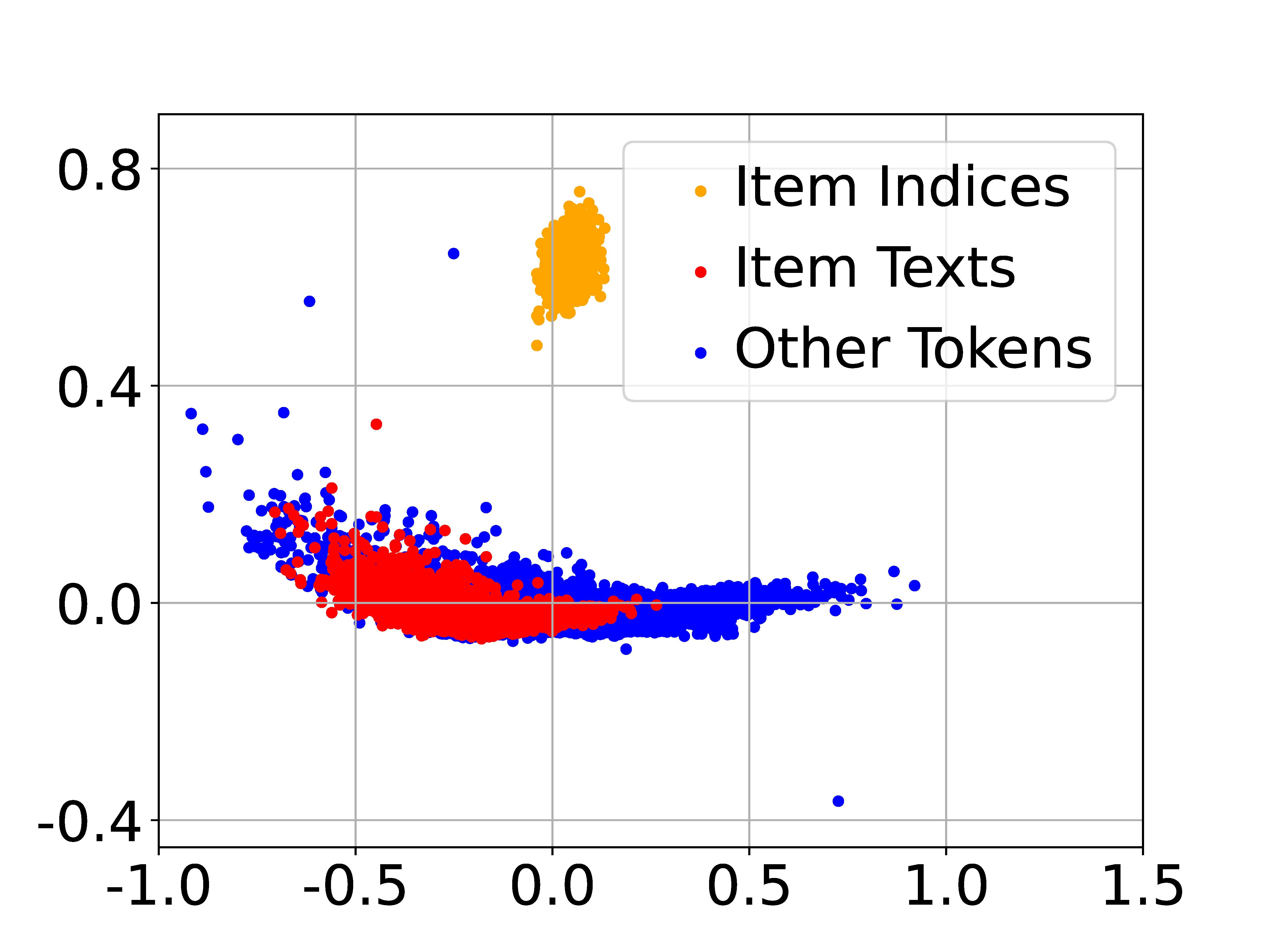}}
\hspace{0.1mm}
\subfigure[LC-Rec that consists of a series of alignment tasks.]{\includegraphics[width=0.485\linewidth]{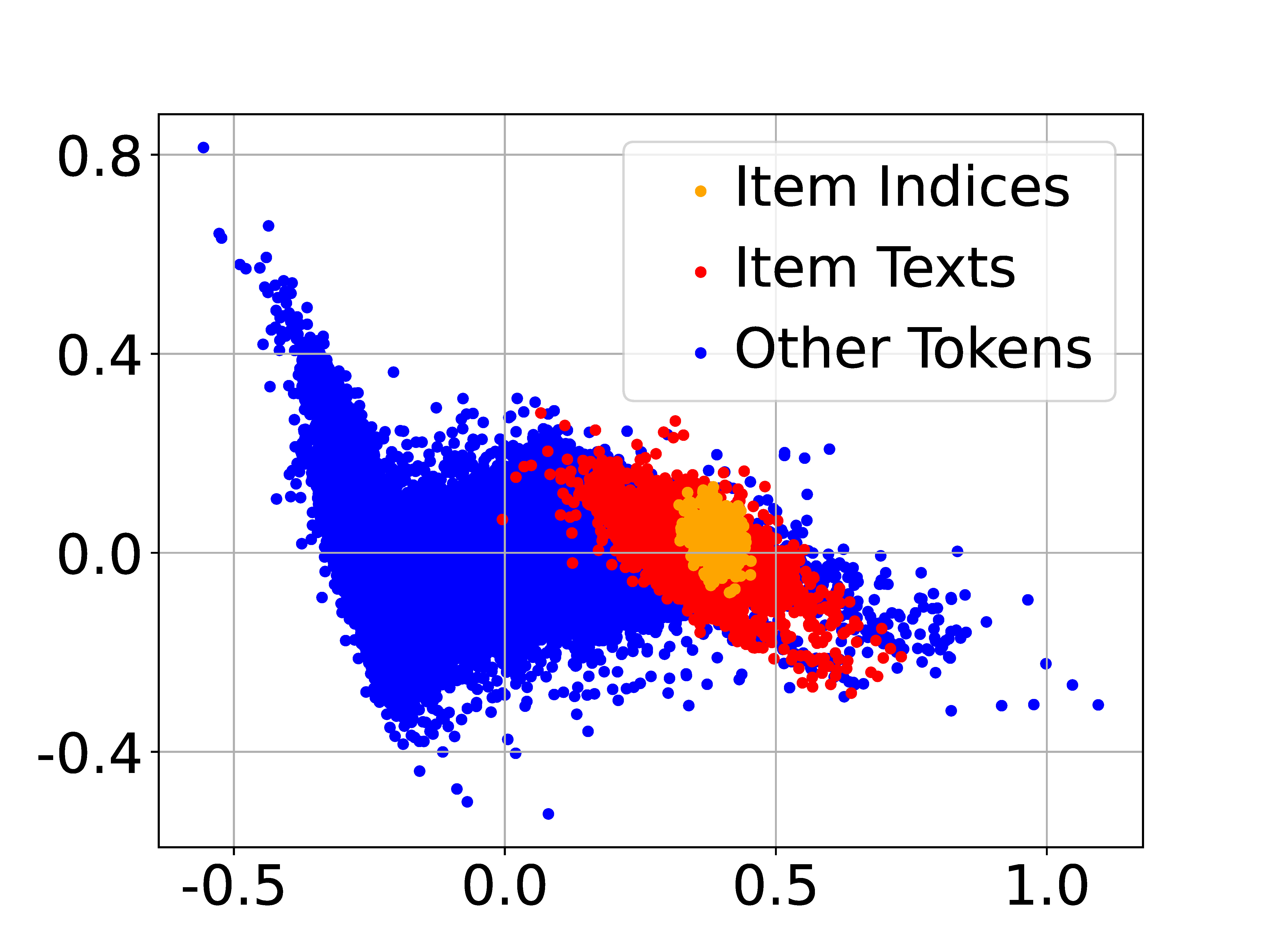}}
\caption{2D visualization of LLM token embeddings via PCA.}
\label{fig:emb-pca}
\end{figure}

\paragraph{Embedding visualization analysis}
To further investigate the effects of our proposed framework in terms of semantic integration, we delve deeper to study the connection between item indices and the original semantic space of the LLM.  
Following previous work~\cite{gao2018representation,wang2019improving}, we employ Principal Component Analysis (PCA) to visualize the embeddings corresponding to different tokens. 
As shown in Figure~\ref{fig:emb-pca}, ``Item Indices'' represents index tokens added to the vocabulary, while ``Item Texts'' represents tokens related to item texts (\eg title and description).
According to 2D visualization results, it is evident that a lack of semantic integration leads to incompatibility between item index tokens and the LLM semantic space. In contrast, our framework is capable of incorporating item indices into the LLM and aligning language and collaborative semantics.

\begin{table}[t]
\centering
\caption{Performance on semantically similar negative items. 
% ``LC-Rec (Title)'' denotes that our approach makes recommendations based on item titles rather than indices.
}
\label{tab:hard_rec}
\renewcommand\arraystretch{1.1}
\setlength{\tabcolsep}{0.9mm}{
\begin{tabular}{l|p{1.9cm}<{\centering}p{2.1cm}<{\centering}p{1.9cm}<{\centering}}
\hline
Methods&  Language Neg. &  Collaborative Neg.& Random Neg.\\
\hline
SASRec&  73.52&  52.25& 89.78\\
LLaMA&  56.67&  51.23& 61.14\\
ChatGPT&  60.94&  51.30& 66.66\\
LC-Rec (Title)&  67.74 &  56.72& 84.64\\
LC-Rec &  75.73&  60.01& 90.19\\
\hline
\end{tabular}}
\end{table}

\paragraph{Performance on semantically similar negative items}
% As mentioned above, we aim to integrate language and collaborative semantics for improving LLMs in recommender systems through our framework. 
% Previous experiments have demonstrated the effectiveness of this semantic integration in performing full ranking across the entire item set. 
% Here, in order to individually analyze the practical significance of these two semantics in recommendations, we evaluate our LC-Rec by selecting negative examples that are similar in either language or collaborative semantics. 

In order to understand why integrating language and collaborative semantics can improve LLMs in recommendation tasks, we further evaluate our LC-Rec with a ranking task with different negative samples that are similar to ground truth in either language or collaborative semantics.
Specifically, we first select two types of semantically similar negative items: (1) Items with similar language semantics, which are selected based on the cosine similarity between item text embeddings. (2) Items with similar collaborative semantics, which are selected based on the cosine similarity between item embeddings from the trained SASRec~\cite{kang2018self} model. 
Subsequently, we use the same test data as sequential recommendation task and utilize the model to choose between the ground-truth target item and the negative item with similar language/collaborative semantics. 
In addition, we use random negative items as a comparison benchmark and measure the performance by accuracy.

We adopt SASRec, LLaMA without fine-tuning, and ChatGPT as the comparison methods. ``LC-Rec (Title)'' refers to our approach but makes recommendations based on item titles rather than indices. The results are shown in Table~\ref{tab:hard_rec}. In the task of distinguishing items with similar language semantics, our method achieved the best performance, benefiting from the integration of collaborative semantics implied by recommender systems. 
% 
% The accuracy results on Games dataset are shown in Table~\ref{tab:hard_rec}, we adopt SASRec~\cite{kang2018self}, LLaMA~\cite{touvron2023llama} without fine-tuning, and ChatGPT as comparison methods. ``LC-Rec (Title)'' refers to our approach but makes recommendations based on item titles rather than indices.
% In the case of language semantic similarity, our method achieved the best performance, benefiting from the integration of collaborative semantics implied by recommender systems.
% 
Additionally, substituting item indices with titles for recommendations also yielded competitive results, which can be attributed to the implicit alignment between item indices and titles within our model.
Another task, distinguishing items with similar collaborative semantics, is often considered more challenging. This is due to the fact that the item with similar collaborative semantics may also have language semantic relevance to the ground-truth target item. 
However, even for such a difficult task, our LC-Rec still shows better performance than the strong baselines, thanks to the unification of language and collaborative semantics.
Furthermore, the non-fine-tuned LLaMA and ChatGPT perform sub-optimally in these challenging scenarios, demonstrating that utilizing LLMs directly for recommendation purposes is often inadequate due to the large gap between recommendation tasks and natural language tasks.

% \begin{figure}[]
% \centering
% \includegraphics[width=0.40\textwidth]{./fig/case1.pdf}
% \caption{Generate the item title based on different numbers of codewords. As the number of codewords increases, the generated content progressively converges towards the target title, and the semantic changes show a trend from coarse to fine.}
% \label{fig:case1}
% \end{figure}

% \begin{figure}[]
% \centering
% \includegraphics[width=0.5\textwidth]{./fig/case2.pdf}
% \caption{Related items are generated based on item indices and similar items are recalled based on text embedding cosine similarity. 
% Compared to those based solely on language semantics, related items generated using item indices that integrate both language and collaborative semantics are more suitable for recommendation scenarios.}
% \label{fig:case2}
% \end{figure}

\begin{figure*}[]
\centering
\subfigure[Generate the item title based on different number of indices.]{\includegraphics[width=0.9\textwidth]{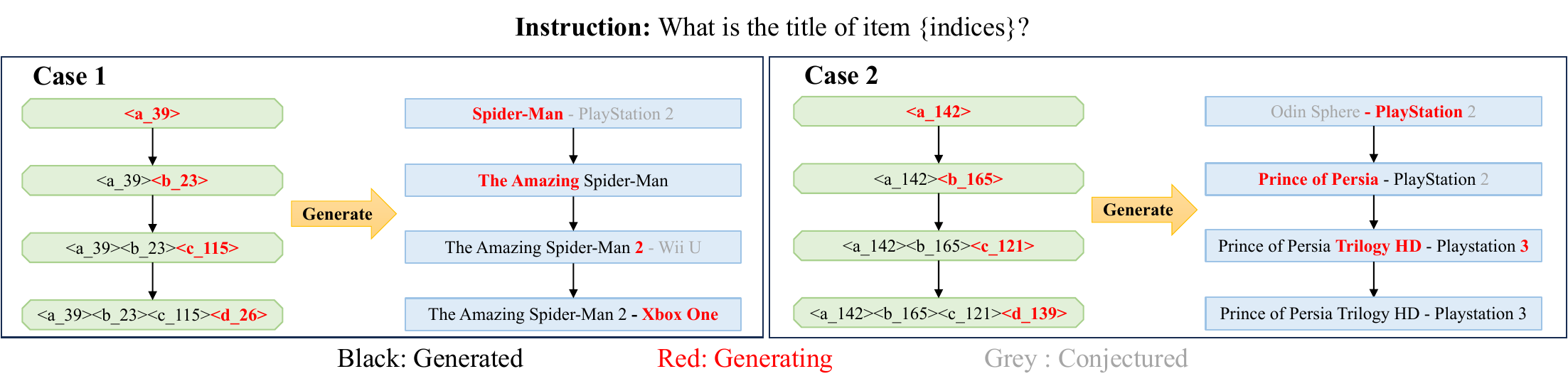} \label{fig:case:gen}}

\subfigure[Related items are generated based on item indices or recalled based on text embedding cosine similarity.]{\includegraphics[width=0.85\textwidth]{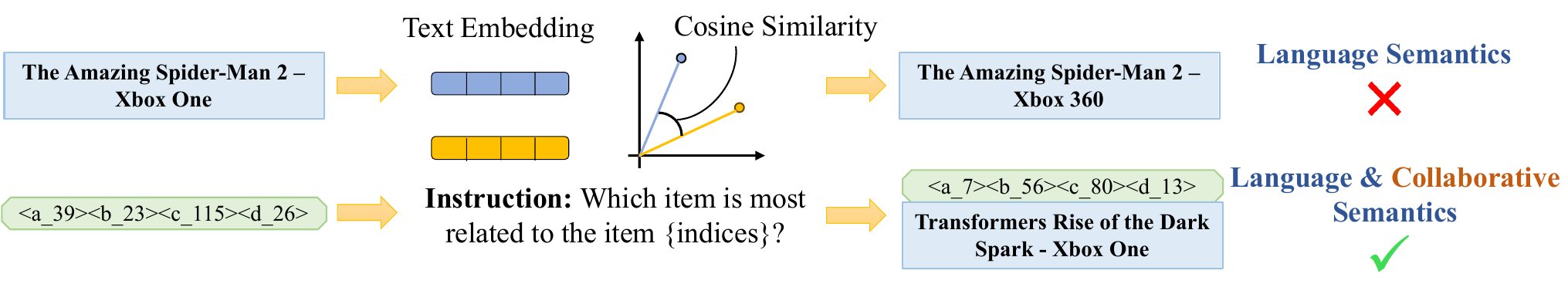}  \label{fig:case:sim}}
\caption{Case study about the semantics within item indices. For the cases in Figure~\ref{fig:case:gen}, it can be observed that as the number of index increases, the generated content progressively converges towards the target title, and the semantic changes show a trend from coarse to fine. For the cases in Figure~\ref{fig:case:sim}, compared to those based solely on language semantics, related items generated using item indices that integrate both language and collaborative semantics are more suitable for recommendation scenarios.}
\label{fig:case}
\end{figure*}

\begin{figure}[]
\centering
\includegraphics[width=0.35\textwidth]{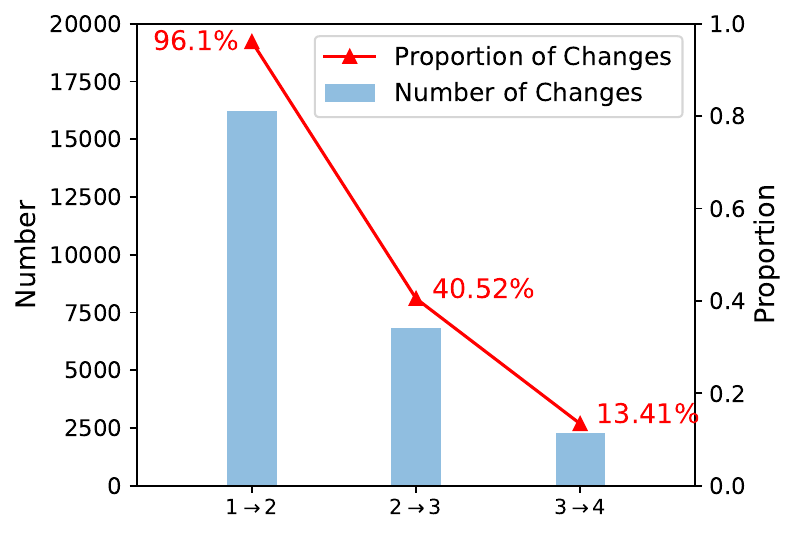}
\caption{Content changes caused by each level index.}
\label{fig:changes}
\end{figure}

\subsection{Case Study}

To intuitively explore the semantic information implicitly learned in the item indices, we present two types of illustrative cases in Figure~\ref{fig:case}.

On the one hand, we analyze the hierarchical semantics in the multi-level item index. Specifically, we initially attempt to generate the item title using only the first index and gradually include more until all four indices are used.
% 
% On the one hand, we generate corresponding item information by utilizing varying numbers of index codewords, allowing us to analyze the semantic information present at different levels within the item index.
% To be specific, we initially attempt to generate the item title using only the first codeword from the item index.
% Then, we progressively incorporate additional codewords, starting with the first two and gradually increasing until all four index codewords are utilized.
% 
As shown in Figure~\ref{fig:case:gen}, when relying solely on the first-level index, the generated content often fails to match the ground-truth item, but it already possesses some relevant semantic information. 
For example, in the first case, a single index can generate the keyword ``Spider-Man'', whereas in the second case, a game belonging to the same categories (\ie adventure) and similar platform  (\ie PlayStation) as the ground-truth item can be generated.
As more indices are included, the generated content progressively converges towards the target title. 
Notably, at the second level, our LC-Rec is already capable of inferring the item name to a significant extent. The subsequent third level further refines the semantic information, while the fourth level contains relatively less semantic information, which is consistent with the coarse-to-fine quantization process employed during index construction. 
Moreover, we also count the proportion of generated results changes caused by each level of indices. As shown in Figure~\ref{fig:changes}, also consistent with our conjecture, the proportion of content changes gradually decreases as the index level increases.

On the other hand, we try to generate the item that are most relevant or similar to a given item through its indices. We then compare the generated results with the similar item obtained based on cosine similarity between item text embeddings.
As presented in Figure~\ref{fig:case:sim}, the similar item generated by our LC-Rec is a game of the same category and platform as the source item, while a duplicate game for another platform is obtained simply based on language semantic similarity. In recommendation scenarios, the former that integrates both language and collaborative semantics is usually more suitable to meet user needs.

% \begin{table}
% \centering
% \caption{Performance on NLP benchmarks.}
% \label{tab:nlp}
% \renewcommand\arraystretch{1.2}
% \setlength{\tabcolsep}{1.2mm}{
% \begin{tabular}{lcccccc}
% \hline
% \multirow{2}{*}{Methods} & \multicolumn{5}{c}{MMLU} & \multirow{2}{*}{BBH} \\
% \cline{2-6}
% & Humanities & STEM & Social Sciences & Other & Average &     \\
% \hline
% LLaMA                                        & 34.0         & 30.6 & 38.2            & 38.3  & 35.2    & 33.25                                    \\
% LC-Rec                                        & 25.1       & 26.9 & 26.8            & 24.9  & 25.8    & 27.36  \\
% \hline
% \end{tabular}}
% \end{table}

% \subsection{Primitive Ability of Large Language Model}

% Our LC-Rec is fine-tuned based on LLaMA~\cite{touvron2023llama}, which ideally equips it with similar natural language processing abilities. In this section, we evaluate our approach on two widely used benchmarks in natural language processing (\ie MMLU~\cite{hendrycks2020measuring} and BBH~\cite{srivastava2022beyond}), and the results are presented in Table~\ref{tab:nlp}.
% As anticipated, the performance of LLM in natural language tasks is somewhat compromised after training on recommendation tasks. This is an aspect that researchers overlook.
% As LLMs continue to scale, striking a balance between achieving excellent recommendation performance and preserving LLM's primitive abilities may be the ideal direction for our future efforts.

\section{Conclusion}
In this paper, we proposed a LLM-based recommendation approach, named LC-Rec. 
In order to adapt LLMs to sequential recommendation tasks, we focused on two main aspects: item indexing and alignment tuning. 
Concretely, we introduced a vector quantization method combined with uniform semantic mapping for item index learning.  
%Besides, the extended index-related parameters will flexibly learn semantic information that is beneficial to recommendations during the subsequent instruction tuning of the LLM.
To facilitate the integration of item indices into the LLM, we proposed a series of semantic alignment tasks to align language and collaborative semantics for recommendation. 
These tasks include sequential item prediction, explicit index-language alignment, and implicit recommendation-oriented alignment. 
Based on the learned item indices, our approach employed these alignment tuning tasks to effectively adapt LLMs for sequential recommendation.   
%Based on the aforementioned approach, our goal is to enable the LLM to grasp the item language and collaborative semantics through its index, thereby allowing the LLM's remarkable semantic comprehension and generation capabilities to be fully utilized in sequential recommendation.
Extensive experiments on three large datasets demonstrated the effectiveness of our approach, outperforming a number of competitive baseline models. 

%Currently, our model is still 
As future work, we will explore how to extend the current approach in a multi-turn chat setting, so that it can support more flexible interaction with users. In addition, we will also investigate how to better reserve the general abilities of LLMs when making domain adaptations. 

%will explore how to maintain the primitive abilities of LLMs while adapting to recommendation tasks. On this basis, we will further try to apply the current method to scenarios such as dialogue recommendation that combines natural language dialogue, user intention understanding, and personalized item recommendation.

\balance

% \section*{Acknowledgment}
% This work was partially supported by National Natural Science Foundation of China under Grant No. 62222215, Beijing Natural Science Foundation under Grant No. 4222027, Fund for Building World-Class Universities (Disciplines) of Renmin University of China, and Beijing Outstanding Young Scientist Program under Grant No.BJJWZYJH012019100020098. Xin Zhao is the corresponding author.

\bibliographystyle{IEEEtran}
\bibliography{ref}

\end{document}